\documentclass[12pt]{article}

\usepackage{amsmath,amssymb,graphicx,cite} 
\usepackage{epsf}
\usepackage{pstricks}

\newcommand{\dslash}[1]{\displaystyle{\not} #1}
\newcommand{\Tr}{\mathrm{Tr~}}

\newcommand{\beq}{\begin{equation}}
\newcommand{\eeq}{\end{equation}}

\begin{document}
\begin{titlepage}

\vskip.5cm
\begin{center}
{\huge \bf  Revealing Randall-Sundrum \\ Hidden Valleys}
\end{center}

\begin{center}
{\bf {Don Bunk}$^a$,  {Jay Hubisz}$^a$} \\
\end{center}
\vskip 8pt

\begin{center}
$^a$ {\it  Department of Physics, Syracuse University, Syracuse, NY  13244}

\vspace*{0.2cm}

{\tt  djbunk@physics.syr.edu, hubisz@physics.syr.edu}
\end{center}

\vglue 0.3truecm

\begin{abstract}
\vskip 3pt \noindent
We study 5D gauge symmetries in the Randall-Sundrum geometry that are hidden from the standard model through either small 5D gauge coupling, or through vanishing quantum numbers for the standard model fields.  Geometric warping of 5D gravity creates a TeV scale bridge from the standard model to the hidden sector gauge fields.   We apply these concepts to a revival of the electroweak axion model, in which the dynamics of Peccei-Quinn symmetry breaking occur at the TeV scale.

\end{abstract}
\end{titlepage}

\newpage

\section{Introduction}
\label{intro}
\setcounter{equation}{0}
\setcounter{footnote}{0}

The Randall-Sundrum mechanism of resolving the hierarchy between the electroweak sector and the Planck scale has now seen numerous implementations~\cite{RS}.  Higgsless models of electroweak symmetry breaking achieve masses for Standard Model (SM) particles through the boundary conditions chosen for bulk gauge fields and fermions~\cite{HL1,HL2,HL3}.  Modern 5D composite Higgs models accomplish electroweak symmetry breaking through a Higgs that originates from a 5-dimensional gauge symmetry, or through brane localized scalar fields~\cite{CH1,CH2}.

These models have been constructed with a certain minimalist approach, introducing the fewest new particles or symmetries to accomplish the goal of reducing the fine-tuning problem in the electroweak sector.  We study cases in which there may be additional light fields which reside within the same RS geometry.   In principle, such fields may be playing an important role in solving other issues within the SM, such as the strong CP problem~\cite{strongCP}, however  we take the approach of studying a generic class of models in which there are new light particles that have greatly suppressed couplings to SM fields.  The most likely candidates for such light particles would be Goldstone bosons, whose masses are small in comparison with the weak scale due to a(n approximate) shift symmetry, or new gauge fields, protected by a 4D gauge symmetry.  Both classes of particles, Goldstone bosons and 4D gauge fields arise naturally from 5D gauge symmetries as a consequence of the different boundary conditions~\cite{orbifoldBC} that one may impose on the 5D gauge transformations (see~\cite{TASI} for reviews and additional references).

A main result of this analysis is the observation that extra-dimensional gravitational excitations~\cite{graviscalars,radcsaba,kkgravphen}, whose couplings to such hidden sectors (HS) is independent of gauge-coupling~\cite{bulkradion}, create a bridge between the visible and hidden fields.  Randall-Sundrum models are thus a natural setting for Hidden Valley models, in which a new sector is separated from the SM through an ``energy-barrier"~\cite{HV1,HV2}.  In RS scenarios, the role of the energy-barrier is played by the extra-dimensional gravitational excitations of the Randall-Sundrum geometry.

As an explicit example, we construct a novel axion solution to the strong CP problem which is in some senses a revival of the earliest axion models where \emph{electroweak scale} physics produces a Peccei-Quinn (PQ) axion~\cite{PQ,WWaxion}.  This 5D axion is hidden by a small extra-dimensional gauge coupling, but has TeV-scale associated Kaluza-Klein excitations, unlike in previous models~\cite{warpedaxions}, in which the IR brane is coincident with the scale of PQ symmetry breaking.  This model shares some features with composite axion models~\cite{compaxion,compaxion1,compaxion2}, although the effective compositeness scale in this case is close to the electroweak scale, and is decoupled from the scale associated with the axion coupling constant.  The gravity sector can act as a bridge to the axion sector, resulting in a greatly altered collider phenomenology, and necessitating a re-evaluation of the usual astrophysical bounds on such light fields.

In Section 2 we describe the basic setup for an RS hidden gauge sector.  In Section 3 we discuss direct couplings of SM fields to the hidden sector.  In section 4, we calculate the couplings of RS gravitational fluctuations to hidden sector fields, and in section 5 we describe a toy model in which the RS hidden sector is responsible for producing an axion which resolves the strong CP problem.  In section 6 we describe the collider phenomenology of such hidden sectors, while in section 6 we discuss astrophysical constraints on light hidden RS Goldstone bosons.  In Appendices A and B, we give Feynman rules for the interactions of hidden sector fields with RS gravity, and describe details concerning gauge fixing.

\section{Basic Setup}
\label{setup}
We work in an RS geometry, using the coordinate convention where the metric is conformally flat:
\begin{equation}
ds^2 = \left( \frac{R}{z} \right)^2 \left[ \eta_{\mu \nu} dx^\mu dx^\nu - dz^2 \right]
\end{equation}
Branes at $z=R,R'$ truncate the extra dimension, with $R \sim M_\text{Pl}^{-1}$, and $R' \sim \text{TeV}^{-1}$.  The electroweak hierarchy problem is alleviated as the cutoff scale for radiatively divergent observables in the low energy theory is lowered to near the TeV scale.  It is presumed new physics comes in near this scale which softens this dependence on the UV scale.   The model is constructed on an $S_1/Z_2$ orbifold in order to obtain the chiral spectrum required to reproduce the SM.  The gauge fields are assumed to propagate in the bulk, and the mechanism of electroweak symmetry breaking is left unspecified, as it is model-dependent.  A TeV brane localized Higgs~\cite{RS}, a Higgsless mechanism~\cite{HL1,HL2,HL3}, or a 5D gauge field Higgs~\cite{CH1,CH2}, or some combination of these ideas could be responsible for the generation of fermion and gauge boson masses.

We gauge a new symmetry (not necessarily abelian) in the bulk of the extra dimension.  The 5D Lagrangian for this gauge symmetry is given by:
\begin{equation}
{\mathcal L}_\text{5D} = -\frac{1}{4} \sqrt{g}\left[ g^{MN} g^{RS} B^a_{MR} B^a_{NS}\right] - \frac{1}{2} \sqrt{g} \left( G^a \right)^2 + \sqrt{g} c^a \frac{\delta G^a}{\delta \beta^b} c^b
\end{equation}
The first term is the usual 5D gauge kinetic term, and the second term is a gauge fixing term which removes 5D kinetic mixing between the $B_\mu$ and $B_5$ fields.  The last term is a ghost Lagrangian that restores unitarity to the gauge-fixed non-abelian theory.  In Appendix B, we provide further discussion of gauge fixing.


To determine the spectrum of the gauge sector, we expand the bulk gauge fields in terms of eigenvalues of the 4D gauge equations of motion:  $B_\mu (x,z) = \epsilon_\mu(p) f(z) e^{i p \cdot x}$.The bulk equations of motion for the 4D vector-field wave functions in this geometry are:
\begin{equation}
\label{eq:gaugeeom}
f'' - \frac{1}{z} f'+M^2 f =0
\end{equation}
and the solutions to this eigenvalue problem are
\begin{equation}
\label{bulksol}
f(z) = z \left( A J_1 (m_n z) + B Y_1(m_n z) \right).
\end{equation}
The coefficients A,B and eigenvalues, $m_n$, are found by choosing and imposing boundary conditions and suitably normalizing the 4D effective fields.

We study two scenarios.  First we take boundary conditions that produce $B_5$ zero modes (5D Goldstone bosons) due to breaking the 5D gauge symmetry twice, once on each brane.  In the other scenario, we assume that the 4D gauge symmetry is unbroken on both branes, with resulting $B_\mu$ zero modes.  We also discuss explicit and spontaneous breaking of these symmetries which would lead to Goldstone (gauge) field masses in each of these models, respectively.
\subsection{Hidden RS Goldstones}

To obtain Goldstone bosons from the 5D gauge symmetry, a subgroup of the gauge symmetry must be broken twice, once at the UV brane, $z=R$, and again at the IR brane, $z=R'$.  The boundary conditions which achieve this, and which satisfy the 5D action principle, are $B^a_\mu |_{z=R,R'} = 0$.  In this section, we assume that the entire gauge gauge group is broken twice in this way, and thus the number of Goldstone bosons is equal to the rank of the original bulk gauge symmetry.  We additionally suppress the internal gauge indices, and take the rank of the coset space (the number of Goldstone bosons) to be $N$.

For the $B_5$, the equation of motion in the gauge we choose is:
\begin{equation}
\Box B_5 - \partial_z \left[ z \partial_z \left( \frac{1}{z} B_5 \right) \right] = 0  
\end{equation}
There is a zero mode solution to this equation where $\Box B_5 = 0$.  In this case, the wave function for the $B_5$ zero-mode is given by $B_5 = B_5 (x) \zeta (z)$ with $\frac{R}{z} \zeta (z) = A_0 + B_0 \log z$.

The boundary condition $B_\mu |_{R,R'} = 0$ ensures that there are no necessary boundary gauge fixing terms, and so the boundary conditions for $B_5$ simply arise from the terms coming from integration by parts of the bulk gauge fixing term.  These impose:
\begin{equation}
\left. \partial_z \left( \frac{1}{z} B_5 \right) \right|_{R,R'} = 0,
\end{equation}
and thus the $B_5$ zero mode takes the following form:
\begin{equation}
B_5^{(0)} (x,z) = \frac{\sqrt{2} g_\text{5D} \sqrt{R}}{\sqrt{R'^2-R^2}}  \frac{z}{R}B_5^{(0)} (x) ,
\end{equation}
where the overall coefficient ensures that the $B_5^{(0)}$ is canonically normalized in the 4D effective theory.

The residual gauge symmetry, after adding the gauge fixing term specified in Appendix B, is given by:
\begin{equation}
\Box \beta - z \partial_z \left( \frac{1}{z} \partial_z \beta \right) = 0,
\end{equation}
implying that there is a residual subgroup which is global from the perspective of the 4D coordinates:  $\beta(z) = \beta_0 + \beta_2 z^2$.

The spectrum of $B_\mu$ modes can be found by imposing the boundary conditions on the solutions to the bulk equations of motion, (\ref{eq:gaugeeom}).  The eigenvalue problem is then:
\begin{equation}
\frac{J_1(m_n R') }{Y_1(m_n R')} = \frac{J_1(m_n R)}{Y_1(m_n R)}
\end{equation}
with approximate solutions $m_n R' = 3.83, 7.02, 10.17, 13.32,...$   The $B_5$ Goldstone bosons and their associated vector KK-modes are hidden from the standard model in one of two ways.  Either the gauge coupling associated with this 5D symmetry is very small, $g_\text{5D} \ll \sqrt{R}$, or the SM does not carry quantum numbers under the new symmetry.

The effective scale of symmetry breaking that this Goldstone boson corresponds to is given by (see also~\cite{warpedaxions}), as we will show explicitly in Section~\ref{sec:addingfermions}:
\begin{equation}
f_\text{eff} = \frac{1}{\sqrt{2} R'} \frac{ \sqrt{R} }{g_\text{5D}},
\label{eq:fff}
\end{equation}
and we will also see that couplings of this Goldstone boson to other light fields transforming under the 5D gauge symmetry are suppressed by this breaking scale.  We note that the scale $f_\text{eff}$ can be parametrically larger than the IR scale, $1/R'$ if the 5D gauge coupling is chosen such that $g_\text{5D} \ll \sqrt{R}$.~\footnote{Such choices may be in conflict with the conjectured bounds on gauge couplings that arise by considering the spectrum of charged Planck scale black hole remnants~\cite{nosmallgaugecouplings}.  While perfectly sound from an effective field theory point of view, it is likely that a new effective cutoff is introduced which is given approximately by $\Lambda = g_\text{5D} \sqrt{R}$, parametrically smaller than the 5D Planck scale.  New physics (perhaps stringy in nature) must appear at this scale which drive the gauge coupling to be strong enough to avoid these bounds.}  


\subsection{Hidden RS Gauge symmetries}

In this section, we briefly analyze the scenario in which the 4D portion of bulk gauge symmetry is completely unbroken, and there are $B_\mu$ zero modes in the theory.  In this case, the boundary conditions are:
\begin{equation}
\left. \partial_z B_\mu \right|_{z=R',R} = 0
\end{equation}
In this scenario, the residual gauge symmetry on the branes corresponds to transformations that are a function only of the 4D coordinates:
$\left. \partial_z \beta \right|_{z=R,R'} = 0$.  In this case, there is a subgroup of the residual gauge transformations where the gauge transformation parameter is a function of the 4D coordinates only:  $\beta = \beta(x)$.  Thus this 5D gauge symmetry has a residual unbroken 4D gauge symmetry corresponding to the $B_\mu$ zero-mode.  The remaining gauge freedom contains $z$-dependence, and corresponds to transformations of the tower of $B_\mu$ Kaluza-Klein modes.

Using the 5D bulk solution in Eq.~(\ref{bulksol}) in coordination with these boundary conditions, the eigenvalue problem is
\begin{equation}
\frac{J_0(m_n R') }{Y_0(m_n R')} = \frac{J_0(m_n R)}{Y_0(m_n R)}
\end{equation}
with approximate solutions $m_n R' = 0, 2.45,5.56, 8.70, 11.84,...$.  The effective gauge coupling for the zero-mode in terms of the geometrical parameters and the 5D gauge coupling is:
\begin{equation}
g_\text{4D} = \frac{g_{5D}}{\sqrt{R \log \frac{R'}{R}}}
\end{equation}


\section{SM Couplings to RS Hidden Sectors}
\label{sec:addingfermions}

Matter fields in the standard model may have couplings to the HS fields which are suppressed by a small extra dimensional gauge coupling.  In this section we discuss the nature of these couplings to an unbroken HS gauge symmetry, and to a HS gauge symmetry which is broken to a global subgroup, producing a light 5D Goldstone-boson.  We work out the case of a 5D fermion coupled to the HS; couplings to fields with different spin can be derived straightforwardly.

The action for a 5D fermion coupled to a HS $U(1)$ with gauge fields $B_M$ is given by:
\begin{equation}
S = \int d^5x \sqrt{g} \left[ \bar{\Psi} i \displaystyle{\empty}{\not} D \Psi + \frac{c}{R} \bar{\Psi} \Psi \right]
\end{equation}
where $D_M$ is the hermitian gauge covariant derivative
\begin{equation}
D_M = \frac{1}{2} \left( \overrightarrow{\partial}_M - \overleftarrow{\partial}_M \right) - i q B_M,
\end{equation}
and $c$ is the 5D bulk Dirac mass in units of the curvature.  The additional terms involving spin connections that can appear in non-trivial geometries vanish with this metric. 
The 5D Dirac fermion can be expanded in terms of KK-modes:
\begin{equation}
\label{eq:psikk}
\Psi = \sum_n \left( \begin{array}{c} g_n(z) \chi_n (x) \\ f_n (z) \bar{\psi}_n (x) \end{array} \right).
\end{equation}
The functions $\chi_n(x)$ and $\bar{\psi}_n(x)$ are solutions to the 4D Dirac equation, each with mass $m_n$, while the wave functions $f_n$ and $g_n$ are solutions to the 5D equations of motion with eigenvalues $m_n$.

We choose boundary conditions for the 5D fermion such that there is a massless mode (e.g. $(++,--)$ boundary conditions, where $-$ refers to Dirichlet boundary conditions).  Depending on the choice of the bulk mass term, $c$, the zero-mode fermion is either localized on the UV brane ($c< 1/2$), or on the IR brane, $c> 1/2$.


\subsection{Fermion Couplings to a $B_5$ zero mode}
In the case that the extra dimensional gauge symmetry is broken on both branes, and there is a massless $B_5$, there is a set of field redefinitions that may be performed that elucidate the Goldstone nature of this field.  This is in close analogy with the standard prescription in 4D theories with spontaneous global symmetry breaking, where a field $\Phi$ may be redefined as $\Phi \rightarrow e^{i \pi/f} \Phi'$, where $\pi$ are the Goldstone degrees of freedom that couple derivatively, and $\Phi'$ contains only the vev $f$, and the radial fluctuations of the field.  Similarly, fermions $\Psi$ which carry charge $q$ under the global symmetry broken by the vev of $\Phi$ can be redefined as $\Psi \rightarrow e^{ i q \pi/v} \Psi'$, where the transformation law for $\Psi'$ is trivial, with the transformation of $\Psi$ being carried by the shift symmetry of the Goldstone boson.

For the fermion field in our discussion, the field redefinition can be taken to be~\cite{fakegaugino}
\begin{equation}
\Psi(z,x) = \exp \left[ i q \int_{z_0}^z dz' B_5(x,z') \right] \Psi'(z,x).
\label{eq:redef}
\end{equation}
The transformation law for $\Psi'$ is then
\begin{equation}
\Psi' (z,x) \rightarrow e^{i q \beta (z_0)} \Psi' (z,x),
\end{equation}
independent of $z$.  The constant $z_0$ is arbitrary, however it can be chosen in a convenient manner that depends on the 5D EWSB breaking model into which this HS is embedded.  Under this redefinition, for an abelian HS, the fermion gauge invariant kinetic term is modified in the following way:
\begin{equation}
\bar{\Psi} i \dslash{D} \Psi \rightarrow \bar{\Psi}' i \dslash{D}_4 \Psi' - \bar{\Psi}' i \gamma^5 \partial_5 \Psi' -q \int_{z_0}^z dz' \partial_\mu B_5(z') \bar{\Psi}' \gamma^\mu \Psi'.
\end{equation}
Note that the $B_5$ now couples derivatively in the 4D coordinates, as expected for a Goldstone boson.  In the presence of additional fields, such as Higgs scalars which carry HS quantum numbers,  (as was the case in~\cite{fakegaugino}), the most convenient redefinition may be slightly different, and could involve the scalar degrees of freedom.  

We can now determine the effective global symmetry breaking scale that produces the $B_5$ Goldstone boson,  and read off its corresponding classically conserved current.   From the action after the redefinition, we see that the interactions of the $B_5$ zero mode with fermions is given by:
\begin{eqnarray}
\mathcal{L}_\text{eff}  &=& - q \int dz \sqrt{g} \int_{z}^{R'} dz' A_0 \left(\frac{z'}{R} \right)\left( \partial_\mu B_5^{(0)} (x) \right) \bar{\Psi}' e_a^\mu \gamma^a \Psi' \nonumber \\ 
&=& -q \partial_\mu B_5^{(0)}(x) \int_R^{R'} dz \frac{g_{5D}}{\sqrt{2 R}} \left( \frac{R}{z} \right)^4 \frac{z^2 - z_0^2}{R'} \left( \bar{\Psi}'  \gamma^\mu \Psi' \right) \nonumber \\
&\equiv& -q \partial_\mu B_5^{(0)} (x) \sum_{m,n} \left[ \frac{1}{f^{mn}_L} \bar{\chi}_m \sigma^\mu \chi_n +  \frac{1}{f^{mn}_R} \psi_m \bar{\sigma}^\mu \bar{\psi}_n  \right]
\label{eq:derint}
\end{eqnarray}
where
\begin{eqnarray}
&&\frac{1}{f^{mn}_L} = \frac{g_{5D}}{\sqrt{2 R}}  \int_R^{R'} dz  \left( \frac{R}{z} \right)^4 \frac{z^2 - z_0^2}{R'} g_m (z) g_n (z)\nonumber \\
\text{and} && \frac{1}{f^{mn}_R} = \frac{g_{5D}}{\sqrt{2 R}}  \int_R^{R'} dz  \left( \frac{R}{z} \right)^4 \frac{z^2 - z_0^2}{R'} f_m (z) f_n(z).
\label{eq:feff}
\end{eqnarray}
The most convenient choice for $z_0$ is model dependent, depending primarily on additional brane localized sources of explicit breaking of the 5D gauge symmetry.  For example, a Dirac-type mass that mixes 2 5D fermions on the IR brane (one producing a LH zero mode, the other a RH zero mode) would transform under the above redefinition as:
\begin{equation}
M \bar{\Psi}_L \Psi_R + \text{h.c.} \rightarrow M \exp \left[ i (q_R - q_L) \int_{z_0}^{R'} dz' B_5 \right] \bar{\Psi}'_L \Psi'_R +\text{h.c.},
\label{eq:massbreaking}
\end{equation}
thus introducing additional interactions of the $B_5$ zero mode with fermions which are physically equivalent to the types of interactions in Eq.~(\ref{eq:derint}).  Such interactions contribute to the amplitudes in such a way as to give the same effective coupling in any physical process.  Choosing $z_0 = R'$ for such a model eliminates this additional contribution to the coupling, such that the entire interaction with fermions can be read from Eq.~(\ref{eq:derint}).

Let us assume that there is a $\chi$ zero mode arising in $\Psi'$, and that there is a bulk Dirac mass term, $c$, that determines the localization of this zero mode.  The zero mode profile is then given by:
\begin{equation}
g_0 (z) = \kappa \left( \frac{z}{R} \right)^{2-c}.
\end{equation}
This fermion is localized towards the UV (IR) brane for $c > (<) 1/2$.  Plugging this wave function into the expressions in Eq.~\ref{eq:feff}, we find that the associated breaking scale for left handed zero mode fermions as a function of the $c$-parameter is given by:
\begin{equation}
f^{00}_L = \left\{ \begin{array}{ll} \frac{1}{R'} \frac{\sqrt{R}}{\sqrt{2} g_5}  & c > 1/2~~~\text{UV localized} \\
 \frac{1}{R'} \frac{\sqrt{R}}{\sqrt{2} g_5} \frac{1}{3/2-c} & c < 1/2~~~\text{IR localized,} \end{array} \right.
 \end{equation}
 roughly confirming the interpretation of the 5D gauge coupling in terms of a symmetry breaking scale, Eq.~(\ref{eq:fff}).
\newline\newline
\subsection{Gauge Field Couplings to a $B_5$ zero mode}

The redefinition~(\ref{eq:redef}) may produce a non-trivial Jacobian in the path integral measure, reflecting explicit breaking of the global shift symmetry of the $B_5$ Goldstone boson through anomalies~\cite{anomalies,fujikawa}.    Such anomalies result in couplings of the $B_5$ zero mode to the 5D gauge fields, including SM gluons and photons~\cite{fakegaugino,warpedaxions}.

In the bulk, the theory is vector-like, and there can be no anomalies, however the boundary conditions are chosen to project out a chirality on the branes to obtain a low energy chiral spectrum.  The contributions of a single 5D fermion with a chiral zero mode to the anomaly are evenly distributed on the boundaries of the space, with half of the chiral anomaly localized on the UV brane, and the other half on the IR brane~\cite{nimanom,warpedanom}.  Under an anomalous 5D gauge transformation, the action shifts by:
\beq
\delta \mathcal{S} = \int d^4x \int_R^{R'} dz~ \beta \partial_M J^M - \int d^4x \left. \beta  J^5 \right|^{R'}_R \equiv  \int d^5x ~ \beta \mathcal{A},
\eeq
with $ J^{M}$ given by
\beq
J^{M} \equiv \sqrt{g} \bar\Psi \gamma^{M} \Psi,
\eeq
and the anomaly, $\mathcal{A}$, is given by:
\begin{eqnarray}
&\mathcal{A}(x,z) = \frac{1}{2} \left[ \delta(z-R) + \delta(z-R') \right]  \sum_f q^f \left( \frac{q_Y^{f2}}{16 \pi^2} F \cdot \tilde{F}+\frac{\Tr \tau^f_a \tau^f_a}{16 \pi^2} W \cdot \tilde{W}+\frac{\Tr t^f_a t^f_a}{16 \pi^2} G \cdot \tilde{G} \right)& \nonumber \\
&\equiv \frac{1}{2} \left[ \delta(z-R) + \delta(z-R') \right] {\mathcal Q} (x,z)&
\end{eqnarray}
Such anomalies are not an indication of a ``sick" theory, since the transformation is only anomalous on the boundaries of the extra dimension, where the 5D gauge symmetry is restricted to be global with respect to the 4D coordinates.

The resulting action after the field redefinition~(\ref{eq:redef}), is augmented by the following term:
\begin{equation}
S_\text{anomaly} =-\frac{1}{2} \int d^4x \left[ \int_{R}^{z_0} dz' B_5 \mathcal{Q}(R,x) - \int_{z_0}^{R'} dz' B_5 \mathcal{Q} (R',x) \right]
\end{equation}

In terms of the zero mode $B_5$, which has the profile given above, these interactions are:
\begin{equation}
\mathcal{L}_\text{anom} = \frac{1}{2} A_0 B_5^{(0)} (x) \left[ \left(z_0^2 -R^2 \right) \mathcal{Q}(R) -\left(R'^2 - z_0^2 \right) \mathcal{Q} (R') \right]
\end{equation}
Defining $\mathcal{Q}^\pm \equiv \mathcal{Q} (R') \pm \mathcal{Q} (R)$, we have,
\begin{equation}
\mathcal{L}_\text{anom} = \frac{1}{2} A_0 B_5^{(0)} (x) \left[ \left( 2 z_0^2 - R^2 -R'^2\right) \mathcal{Q}^+ - \left(R'^2 - R^2 \right) \mathcal{Q}^- \right].
\label{eq:anomterm}
\end{equation}

As mentioned above, the physics is not dependent on the choice of $z_0$, however there are choices which are more convenient than others.  Again, in the presence of a Dirac mass term on the IR brane, a sensible choice is $z_0 = R'$.  If another value is chosen, the interactions of the Goldstone boson with fermions arising in equation~(\ref{eq:massbreaking}) will lead to additional triangle loop diagrams which contribute to the interaction in Eq.~(\ref{eq:anomterm}) in such a way as to render the physical result independent of $z_0$.
The anomaly interaction with the choice $z_0 = R'$ is then given by:
\begin{equation}
\mathcal{L}_\text{anom} = \frac{1}{2} A_0 \left(R'^2 - R^2 \right) B_5^{(0)} (x) \left[\mathcal{Q}^+ -  \mathcal{Q}^- \right]
\end{equation}
Finally, plugging in the normalization coefficient for the $B_5$ zero mode, the effective interaction of the $B_5$ zero mode is given by:
\begin{equation}
\mathcal{L}_\text{anom} = \frac{1}{\sqrt{2}} \frac{g_5}{\sqrt{R}}  \sqrt{R'^2 - R^2}  B_5^{(0)} (x) \left[ \mathcal{Q}^+ -\mathcal{Q}^- \right]
\end{equation}
And the effective suppression scale for the anomalous interactions of the $B_5$ zero mode with SM gauge bosons is approximately
\begin{equation}
f_\text{anom}^{00} = \frac{1}{R'} \frac{\sqrt{R}}{\sqrt{2} g_5},
\end{equation}
in agreement with the effective Goldstone boson scale arising from the couplings to fermion zero-modes in Section~\ref{sec:addingfermions}.  There are additional interactions of the $B_5$ with gauge boson KK-modes when the anomalies $\mathcal{Q}^\pm$ are expressed in terms of a KK-mode expansion.


\section{Couplings to RS Gravity}

Unlike the couplings of the Goldstone sector to SM fields, the couplings of the excitations of RS gravity to the gauge fields $B_\mu$ (or physical $B_5$ Goldstone bosons) are independent of the 5D gauge coupling~\cite{kkgravphen,bulkradion}.  Thus while the gauge sector may be ``hidden" from the SM fields, the couplings of the hidden sector to TeV brane localized gravitational waves are suppressed only by the IR brane local cutoff scale.  In this section, we calculate the couplings of RS gravitational excitations (the radion and the first two tensor modes) to the hidden sector gauge fields.

We begin by reviewing the KK-reduction of the 5D metric including linearized fluctuations.  The usual Einstein-Hilbert action is given by:
\begin{equation}
S_{\text{EH}} = -\kappa^2 R^3 \int_R^{R'} dz\int d^4 x \sqrt{g} \left(  \mathcal{R} - \Lambda \right)
\end{equation}
The distance element on this space, including linearized perturbations which solve the vacuum Einstein equations, is given by:
\begin{equation}
ds^2 = \left(\frac{R}{z}\right)^2 \left[ e^{-2 F(z,x)} \eta_{\mu\nu} dx^\mu dx^\nu + h_{\mu\nu} dx^\mu dx^\nu - \left(1+2 F(z,x)\right)^2 dz^2 \right],
\label{eq:distelement}
\end{equation}
where $h_{\mu \nu}$ is transverse and traceless, and contains the 4D graviton plus Kaluza-Klein excitations.  $F$ is the radion field, expressed after canonical normalization as
\begin{equation}
F(z,x) = \left( \frac{z}{R'} \right)^2 \frac{ r(x) }{\kappa \Lambda_r}
\end{equation}
Plugging this radion excitation into the above EH action shows that the normalization factor which sets the scale of the radion coupling to other fields is given by $\Lambda_r = \sqrt{6}/R'$.

The transverse traceless perturbations, $\tilde{h}_{\mu\nu} \equiv \left( R/z \right)^2 h_{\mu\nu}$ satisfy the following bulk equation of motion:
\begin{equation}
\tilde{h}''_{\mu\nu} + \frac{1}{z} \tilde{h}'_{\mu\nu}-\frac{4}{z^2} \tilde{h}_{\mu\nu}- \Box \tilde{h}_{\mu\nu} =0
\end{equation}
while the boundary conditions require 
\begin{equation}
\left(z^2 \tilde{h}_{\mu\nu} \right)' |_{R,R'} = 0
\end{equation}
After imposing the boundary condition at $z=R$, with the ansatz $\tilde{h}_{\mu\nu} = \sum_n \phi_n (z) \frac{\hat{h}_n(x)_{\mu\nu}}{\kappa \Lambda_n}$, the KK-graviton wave functions are given by:
\begin{equation}
\phi_n(z) = \left( \frac{R}{R'} \right)^2 \left[ J_2(m_n z) - \frac{J_1 (m_n R)}{Y_1 (m_n R)} Y_2(m_n z) \right].
\end{equation}
Note that we have given the 4D modes $h_n(x)$ mass dimension 1, associating a scale with the couplings of each graviton KK-mode that is calculated by imposing canonical normalization on the 4D modes.  The prefactor $\left( R/R' \right)^2$ is inserted to render the $\Lambda_n$'s sensitive only to the IR scale (where the lower level KK-gravitons are localized).  The scales $\Lambda_n$ are determined by expanding the EH action to quadratic order in the fluctuations, reading off the coefficient of the kinetic terms and enforcing the low energy theory to reproduce the Fierz-Pauli spin-2 kinetic term.   This leads to the following equation for $\Lambda_n$:
\begin{equation}
\frac{1}{R^3} \int dz \left( \frac{z}{R} \right) \phi_n^2 = \Lambda_n^2,
\end{equation}
From which we find $\Lambda_1 R' = .285$, $\Lambda_2 R' = .212$.

The final boundary condition at $z=R'$ determines the solutions to the eigenvalue problem for $m_n$:
\begin{equation}
\frac{J_1(m_n R')}{Y_1(m_n R')} =\frac{J_1(m_n R)}{Y_1(m_n R)}
\end{equation}
This is actually identical in form to the eigenvalue equation for the vector KK-modes of the 5D Goldstone boson in this model, and thus the KK-gravitons have a spectrum identical to the vector KK-modes associated with the Goldstone bosons.

We now calculate the interactions of the radion and KK-gravitons with the light HS fields and the HS KK-modes.  The gravitational excitations couple to the matter stress-energy tensor:
\begin{equation}
S_\text{grav} =  - \frac{1}{2} \int_R^{R'} dz \int d^4x \sqrt{g} (\Delta g)_{MN} T^{MN}
\end{equation}
where the fluctuations including the radion, the graviton, and the KK modes of the graviton are contained in $(\Delta g)_{MN}$.  Using Eq.~(\ref{eq:distelement}) for the distance element, one can read off the interactions of the radion with matter:
\begin{equation}
S_\text{radion} = - \int_R^{R'} dz d^4x \sqrt{g} F(x,z) \left[ \Tr T_{MN} - 3 T_{55} g^{55} \right]
\label{eq:radints}
\end{equation}
while for the graviton and its KK-modes, we have
\begin{equation}
S_\text{grav} =  - \frac{1}{2} \int_R^{R'} dz \int d^4x \sqrt{g} \tilde{h}_{\mu\nu} T^{\mu \nu}
\end{equation}
where the Greek indices are limited to the 4D uncompactified directions.

For a gauge theory, the Maxwell stress-energy tensor (before adding gauge fixing terms) is given by:
\begin{equation}
T_{MN} = \frac{1}{4} g_{MN} B_{RS} B^{RS} - B_{MR} B_{NS} g^{RS}.
\end{equation}
Using the ansatz given above for the $\tilde{h}_{\mu\nu}$ fluctuations, interactions of KK-gravitons with the HS are given by:
\begin{eqnarray}
&&-\frac{1}{2 \Lambda_n} \int d^4x \hat{h}^n_{\mu\nu} \int_R^{R'} dz \sqrt{g} \phi_n(z) T^{\mu\nu} \nonumber \\
 &&= \frac{1}{2 \Lambda_n} \int d^4x \hat{h}^n_{\mu\nu} \int_R^{R'} dz\left( \frac{z}{R} \right) \phi_n(z) \left[ B_{\rho \kappa} B_{\sigma \lambda} \eta^{\kappa \lambda} - B_{\rho 5} B_{\sigma 5} \right] \eta^{\mu \rho} \eta^{\nu \sigma},
\end{eqnarray}
Similarly, plugging the normalized radion field into Eq.~(\ref{eq:radints}), the radion couples in the following way to the HS:
\begin{equation}
\frac{r(x)}{\Lambda_r} \int_R^{R'} dz \left( \frac{z}{R} \right) \left( \frac{R}{R'} \right)^2 \left[ \frac{1}{2} B_{\mu\nu}B_{\rho\sigma}\eta^{\mu \rho} \eta^{\nu \sigma} + 2 \eta^{\mu\nu} B_{\mu 5} B_{\nu 5} \right].
\end{equation}

Using the expressions for the normalized $B_5$, $B_\mu^{(1)}$, and $\hat{h}_{\mu\nu}^{(n)}$, we find the effective 4D Lagrangian coefficients which are summarized in Appendix A.  The couplings are expressed in terms of the normalization factors $\Lambda_n$, the hierarchy between the Planck scale and the position of the UV brane, $\kappa$, and wave function overlap integrals of the $n$'th graviton KK-mode with the HS field, parametrized as $\lambda_{nXX}$, where $X$ are fields residing in the HS.  These coupling constants are robust under variation in the values of $R$ and $R'$, as long as $R' \gg R$.

Note that for a completely brane localized field, $X$, the coupling ratios $\lambda_{n XX}/\Lambda_n \rightarrow \sqrt{2} R'$, bringing our result into agreement with previous publications which have taken the SM fields to be completely localized on the IR brane~\cite{kkgravphen}. 

The primary process which contributes to production is gluon fusion.  The 4D effective Lagrangian  for the couplings of the KK-gravitons to gluons are given by (at tree level)~\cite{kkgravphen}:
\begin{equation}
\mathcal{L}_\text{glue} = \hat{h}^{\mu\nu}_{(1)} G_{\mu \rho} G_\nu^\rho \frac{0.191}{\Lambda_1 \log R'/R} +\hat{h}^{\mu\nu}_{(2)} G_{\mu \rho} G_\nu^\rho \frac{0.028}{\Lambda_2 \log R'/R},
\end{equation}
and the KK-graviton propagator is given by:
\begin{equation}
D_{(n)}^{\mu \nu, \rho \sigma} = \left[ G^{\mu \rho}_{(n)}  G^{\nu \sigma}_{(n)}  + G^{\mu \sigma}_{(n)}  G^{\nu \rho}_{(n)}  - \frac{2}{3} G^{\mu \nu}_{(n)}  G^{\rho \sigma}_{(n)}  \right] \frac{1}{2 \left( k^2 - m_n^2 \right)} \ \ \ \ \ G^{\mu \nu}_{(n)} \equiv \eta^{\mu\nu} - \frac{k^\mu k^\nu}{m_n^2}.
\end{equation}


\section{A TeV-Scale Axion}

In this section, we describe a toy axion model that resolves the strong CP problem and in which a PQ global symmetry is broken at the TeV scale (on the IR brane).  We gauge a $U(1)_\text{PQ}$ symmetry which is broken by boundary conditions on both branes.  The resulting $B_5$ zero mode plays the role of the axion.

In this model the axion is hidden (and its mass supressed) by taking the 5D gauge coupling to be small.  The direct interactions with SM fields are all suppressed by the small extra-dimensional gauge coupling, and with the relation given in Eq.~(\ref{eq:feff}), we deduce that the effective PQ scale is given by:
\begin{equation}
f_\text{PQ} = \frac{1}{R'} \frac{\sqrt{R}}{\sqrt{2} g_5}
\end{equation}
This is the inverse coupling constant that appears in axion interactions that also appear in standard 4D axion models.  For instance, the coupling of the axion to photons and gluons from anomalies is given by
\begin{equation}
c_\text{EM} \frac{B_5}{f_\text{PQ}} F \cdot \tilde{F} + c_\text{QCD} \frac{B_5}{f_\text{PQ}} G \cdot \tilde{G}
\label{eq:axionint}
\end{equation}
where $F$, $G$ and the tildas are the electromagnetic/ gluonic field strengths and their duals.  $c_\text{EM}$ and $c_\text{QCD}$ are the anomaly coefficients.   Below the QCD confinement scale, the second term in Eq.~(\ref{eq:axionint}) leads to an axion mass through instanton effects.  This mass is given approximately by~\cite{WWaxion}
\begin{equation}
m_{B_5}^2 \approx \frac{\Lambda_\text{QCD}^4}{f_\text{PQ}^2}.
\end{equation}
Standard constraints on $f_\text{PQ}$ apply, and the allowed ranges of $f_\text{PQ}$~\cite{PDG} are roughly $10^9~\text{GeV} < f_\text{PQ} < 10^{12}~\text{GeV}$, where the lower bound arises from constraints on supernova cooling rates and the upper bound arises from constraints on the relic abundance of coherent axion oscillations (assuming an order one displacement of the axion field from the CP conserving minimum in the early universe).

Charge assignments under the $U(1)_\text{PQ}$ symmetry are model dependent.  For instance, one could create a hadronic axion model, in which the SM fermions are uncharged, but in which new heavy fermions carrying $SU(3)_C$ charge contribute to the anomaly, and lead to an axion mass.  Another option is to model this 5D axion in a manner similar to the DFSZ axion in terms of the charge assignments:
\begin{equation}
\label{tab:charges}
\begin{array}{|c|c|c|c|c|c|c|c|}\hline
  \     & (H_u) & (H_d) & Q   & \bar{u} & \bar{d} & L         & \bar{e} \\ \hline
Y    & 1/2 & -1/2 & 1/6 & -2/3     &  1/3       &  -1/2  & 1             \\ \hline
\mathrm{PQ}  & 1 & 1  &  -1/2&-1/2   & -1/2      & -1/2   & -1/2         \\ \hline
\end{array}
\end{equation}
The Higgs fields are placed in parentheses as they are not crucial in extra dimensional theories such as Higgless models of electroweak symmetry breaking.   The simplest model in terms of particle content is a Higgsless model augmented by a $U(1)_\text{PQ}$.  The choice of fermion quantum numbers determines the anomaly coefficients $c_\text{EM}$ and $c_\text{QCD}$ in Eq.~(\ref{eq:axionint}), and the most convenient fermion redefinition for a Higgsless theory is given in Eq.~(\ref{eq:redef}), with the choice $z_0 = R'$.

This type of axion model has a strong benefit over previous constructions.  This feature concerns explicit global symmetry breaking terms arising from Planck scale physics which must be suppressed in order to preserve the Goldstone nature of the axion~\cite{symmetrybreaking,compaxion2}.  Without some mechanism to forbid or suppress such operators, non-derivative potential terms for the axion arise and displace the axion from the CP conserving minima of the instanton potential.  In the extra-dimensional construction, such operators, in the 4D effective theory, take the form:
\begin{equation}
\frac{a}{f} \partial_\mu j^\mu_\text{PQ} = \frac{a}{f} \left[ \frac{ g_n }{M_\text{Pl}^n} \mathcal{O}^{4+n} +c_{\text{QCD}}  G \cdot \tilde{G} \right].
\end{equation}
We have also included the term that generates the axion potential from instantons for comparison.  To not spoil the strong CP solution, we must have:
\begin{equation}
10^{-10} c_\text{QCD} \langle G \cdot \tilde{G} \rangle \gtrsim \frac{g_n}{M_\text{Pl}^n} \langle \mathcal{O}^{n+4} \rangle
\end{equation}
With $c_\text{QCD} \langle G \cdot \tilde{G} \rangle \sim \Lambda_\text{QCD}^4$,
this becomes
\begin{equation}
g_n \lesssim 10^{-10} \left(\frac{\Lambda_\text{QCD}}{\mu}\right)^4 \left(\frac{M_\text{Pl}}{\mu} \right)^n
\end{equation}
where $\mu$ is the scale associated with fields appearing in the operator $\mathcal{O}^{4+n}$.  For dimension 5, 6, 7 operators ($n=1,~2,~3$), the scales $\mu$ which satisfy this bound (assuming $g_n = 1$) are $\mu \lesssim 4,~1 \cdot 10^4,~1.4 \cdot 10^6$~GeV.  In this extra-dimensional construction, the terms which correspond to spontaneous symmetry breaking reside on the IR brane, and are naturally of order $TeV$.  Thus the scale $\mu$ is expected to be of order TeV, and even at dimension 6 such operators are not dangerous, a significant improvement on earlier models, in which $\mu$ was tied to the scale $f_\text{PQ}$~\cite{compaxion2,symmetrybreaking}.

Irrespective of the gauge coupling, as shown in the previous section, the RS gravity sector bridges between the SM and this axion sector.  There are thus operators which are suppressed only by the TeV scale associated with the IR brane which connect the SM with the HS axion and its excitations.  In the next sections, we discuss the phenomenology of such hidden sectors, with much of the discussion there being relevant for this axion scenario.


\section{Collider Phenomenology}

Even with greatly suppressed direct couplings, the interactions of the HS with RS-gravity provide a link to SM fields through processes which involve exchange of radions or KK-gravitons.  Observation of gravitational resonances are signals has been considered a smoking gun for extra-dimensional models, so it is vitally important to identify how their phenomenology is modified in the presence of these hidden sectors.  The most dramatic feature involves decays of the radion and KK-gravitons to HS fields, although direct production of HS fields is also possible.


\subsection{Radion and KK-graviton decays to 5D Goldstone Bosons}

Through the interactions shown in Table~\ref{tab:gravcoupGS}, the radion and the graviton KK-modes can decay to the light $B_5$ Goldstone bosons.  These Goldstone bosons may escape the detector, or decay back to light SM states, depending on the model chosen.   In this section, we calculate the partial widths of the radion and KK-gravitons to the light Goldstones.

The radion partial width to Goldstones is given by
\begin{equation}
\Gamma(r \rightarrow B_5 B_5) = \frac{1}{32 \pi} \frac{m_r^3}{\kappa^2 \Lambda_r^2}
\end{equation}
where $\Lambda_r = \sqrt{6}/R'$, independent of the 5D gauge coupling associated with the HS. 

For light radions, where the decay mode $W$'s and $Z$'s is closed, this decay dominates the width, and notably suppresses the $r \rightarrow \gamma\gamma$ branching fraction by roughly a factor of $10$ for radion masses between $114$~GeV and $160$~GeV .  As $\gamma\gamma$ was the most promising channel in which to search for radions at the LHC~\cite{graviscalars,bulkradion}, this is a significant modification in the phenomenology.   The $B_5$ Goldstone bosons produced in these decays stream through the detector since the Goldstones are only very weakly coupled to SM fields.

The radion may also mix with the a Higgs in extra dimensional models that contain a scalar Higgs particle (see e.g.~\cite{graviscalars}).  In this case, the Higgs may have a substantial invisible branching fraction to these Goldstone bosons, even as much as $50$~\% if the relative splitting of the scalar states is comparable to $v R'$.  The amount of mixing however is very model dependent (there may not even be a scalar Higgs particle in the spectrum), and we leave this area as an avenue for future study.  

For heavier radions, where the decays $r \rightarrow W^+ W^-$ contribute to the width, the branching ratio saturates at a value of roughly $20\%$.  

The KK-graviton partial widths to Goldstones are
\begin{equation}
\Gamma(h^{\mu\nu}_n \rightarrow B_5 B_5) = \frac{\lambda_{nB_5 B_5}^2}{1920 \pi} \frac{m_n^3}{\kappa^2 \Lambda_n^2}
\end{equation}
Where $\lambda_{1 B_5 B_5} = -.219$, and  $\lambda_{1 B_5 B_5} = .049$, as can be read from Table~\ref{tab:gravcoupGS}.
This is in agreement with expectations from the Goldstone theorem that this width should be equal to the width to $Z$'s, the Higgs, and half the width to $W$'s, which have been reported in~\cite{kkgravphen}.
The Goldstone equivalence theorem can be then be used to obtain the branching fractions of the KK-graviton to light Goldstones,
\begin{equation}
BR(h^1_{\mu\nu} \rightarrow B^a_5 B^a_5) = \frac{N}{\frac{\Gamma_{\text{top}}}{\Gamma_Z} + 4 + N}
\end{equation}
where $N$ is the number of 5D Goldstone bosons. 
We have neglected the contributions of light UV brane localized fermions and to KK-tops to the total width, as these are typically much smaller~\cite{kkgravphen}.   The branching fraction to a single $U(1)$ Goldstone boson is typically $\mathcal{O}(10 \%)$ for reasonable values of the top-right quark localization parameter, which is the primary variable which determines the ratio $\Gamma_\text{top}/\Gamma_Z$.


\subsection{Radion and KK-graviton decays to hidden 5D gauge fields}

The radion width to gauge boson zero modes is given by:
\begin{equation}
\Gamma(r \rightarrow B_\mu B_\mu ) = \frac{m_r^3}{128 \pi \kappa^2 \Lambda_r^2 \log^2 R'/R}
\end{equation}
Again, since these light vector modes are assumed to couple only very weakly (or not at all) with SM particles, these particles would manifest as missing energy at colliders.  Unlike the Goldstone $B_5$ HS, these invisible decays only contribute modestly to the total width of the radion, and are of roughly the same size as the branching fraction to $\gamma\gamma$.  Thus the radion phenomenology is not greatly altered.  The smaller branching fraction relative to the Goldstone HS scenario is due to the extra log suppression in the couplings of the radion to the flat profile of the $B_\mu$ zero modes.

The level-1 KK-graviton width to gauge boson zero modes is given by
\begin{equation}
\Gamma( h_{\mu\nu}^{(1)} \rightarrow B_\mu B_\mu ) = \left( \frac{.191}{\kappa \Lambda_1 \log R'/R } \right)^2 \frac{m_{(1)}^3}{1536 \pi}
\end{equation} 
and exhibits the same log suppression as the radion decays.  Thus the branching fraction to $B_\mu$ modes will be very small compared to the fractions to SM massive gauge fields, and similar with the branching fraction to photons (in fact the branching fractions are identical up to loop corrections).


\subsection{Non-exact shift or gauge symmetries}

The symmetries (HS shift/gauge symmetries) can not be exact/unbroken for most choices of the 5D gauge coupling since there are stringent constraints from astrophysics on new massless scalar fields and long range forces.   The scalars or vector fields in the HS must have some mass.  If the light HS is hidden through a small 5D gauge coupling, and the SM fields have non-vanishing quantum numbers under the gauge symmetry, the Goldstone bosons will decay to SM particles if the HS fields are massive enough.  Depending on how small the extra dimensional gauge coupling is, and the masses of the light pseudo-Goldstone fields, their decays may range from prompt to cosmological time scales.

For the lightest range of HS scalar masses, the 5D Goldstone boson may decay to SM fermions.  The decay width of a light 5D pseudo-Goldstone boson to SM fermions is given by:
\begin{equation}
\Gamma (B_5 \rightarrow \bar{f} f )= \frac{q^2}{4 \pi}\left( \frac{m_f}{f_\text{eff}} \right)^2 m_{B_5}
\end{equation}

The distance traveled by a pseudo-Goldstone boson that couples universally to leptons before decaying to muons (presuming the $B_5$'s have mass less than $2 m_\tau$, and assuming a 5D gauge symmetry charge of $q=1$), is given by:
\begin{equation}
\Delta x = 58 \text{cm} \left( \frac{f_\text{eff}}{10^6 \text{GeV}} \right)^2 \left( \frac{10 \text{GeV}}{m_{B_5}} \right) \sqrt{\left( \frac{ E }{m_{B_5}} \right)^2 - 1 }
\end{equation}
The pseudo-Goldstone modes may also couple to SM quark fields, in which case there will be displaced hadronic decays.

Thus these RS Hidden sectors are a concrete example of a ``Hidden Valley" model~\cite{HV1,HV2}, in which HS fields may be produced at colliders through on-shell production of RS gravitations resonances which subsequently decay into the HS.  The final decay products of the HS fields may be substantially displaced from the production vertex, depending on the choice of the extra dimensional gauge coupling.  Searches have been performed at the Tevatron, with null results thus far~\cite{D0search,CDFsearch}

In the case of light HS vector fields, the width to fermions is given in the massless fermion limit by:
\begin{equation}
\Gamma (B_\mu \rightarrow \bar{f} f) = \frac{g^2 m_{B_\mu}}{4 \pi}
\end{equation}
and the decay length in the detector is given by:
\begin{equation}
\Delta x = 20 \text{cm} \left( \frac{10^{-8}}{g} \right)^2 \left( \frac{10 \text{GeV}}{m_{B_\mu}} \right) \sqrt{ \left( \frac{E}{m_{B_\mu}} \right)^2 -1}
\end{equation}


\subsection{Hidden KK-modes at Colliders}

It is also possible that higher level KK-modes of the light HS will be directly produced by collider experiments such as the LHC.  The most likely channel for HS KK-mode production in the light Goldstone scenario is a level one KK-mode in association with the light Goldstone boson:  $gg \rightarrow r (\hat{h}_{\mu\nu}^{(1)} )\rightarrow B_\mu^{(1)} B_5$, where the exchanged particle is either a radion or a level one KK-mode graviton.  For a HS with a residual gauge symmetry it is $gg \rightarrow r  (\hat{h}_{\mu\nu}^{(1)} )\rightarrow B_\mu^{(1)} B_\mu^{(0)}$.

The production cross sections for these processes are very small for two reasons.  Firstly, the KK-modes of the gauge fields are quite massive.  The lowest they could be is in the $2$~TeV range for a typical Higgsless model.   Secondly, the rate is suppressed as RS gravity couplings all come with the normalization factors $\Lambda_r$, or $\Lambda_n$ which are in the TeV range.

For a model with a HS Goldstone boson, taking $R' = (500~\text{GeV})^{-1}$, the LHC cross section at design energy ($14$~TeV CM energy) is $\sigma (gg \rightarrow r \rightarrow B_\mu^{(1)} B_5 ) \approx 1 \cdot 10^{-5}$~pb.   For a model with a HS light gauge field, for the same parameters, the cross section is  $\sigma (gg \rightarrow r \rightarrow B_\mu^{(1)} B_\mu^{(0)} ) \approx 5 \cdot 10^{-6}$~pb.  These are up against the design goals of the LHC, however with high luminosity ($100's$ of $\text{fb}^{-1}$) a few events may be possible.  The HS KK-modes dominantly decay via the channel $B^{(1)}_\mu \rightarrow B_5 r (B^{(0)}_\mu r )$ for HS Goldstone (gauge) field.  If the light HS Goldstone (gauge) field  can decay to SM leptons within the detector, there is hope of triggering on and reconstructing even a few such events.


\section{Astrophysical Constraints on RS Goldstone Bosons}

At low energies, the couplings of the hidden sector to RS gravity induce higher dimensional operators involving SM fields that are suppressed only by the TeV scale.  In this section, we calculate the effective operators relevant for main sequence star cooling, and supernova energy loss (see~\cite{astroED} for a related study).  We take into account the contributions from radion exchange, however we neglect the contributions of KK-gravitons, as these are negligible in comparison.  We leave a study of the astrophysical constraints on light RS gauge fields for future work, although we provide expressions for the relevant higher dimensional operators in this section.  Existing light scalar search experiments are not sensitive to the operators that arise from integrating out the RS gravitational excitations.

\subsection{Higher dimensional operators}
\begin{figure}[t]
   \centering
   \includegraphics[width=2.5in]{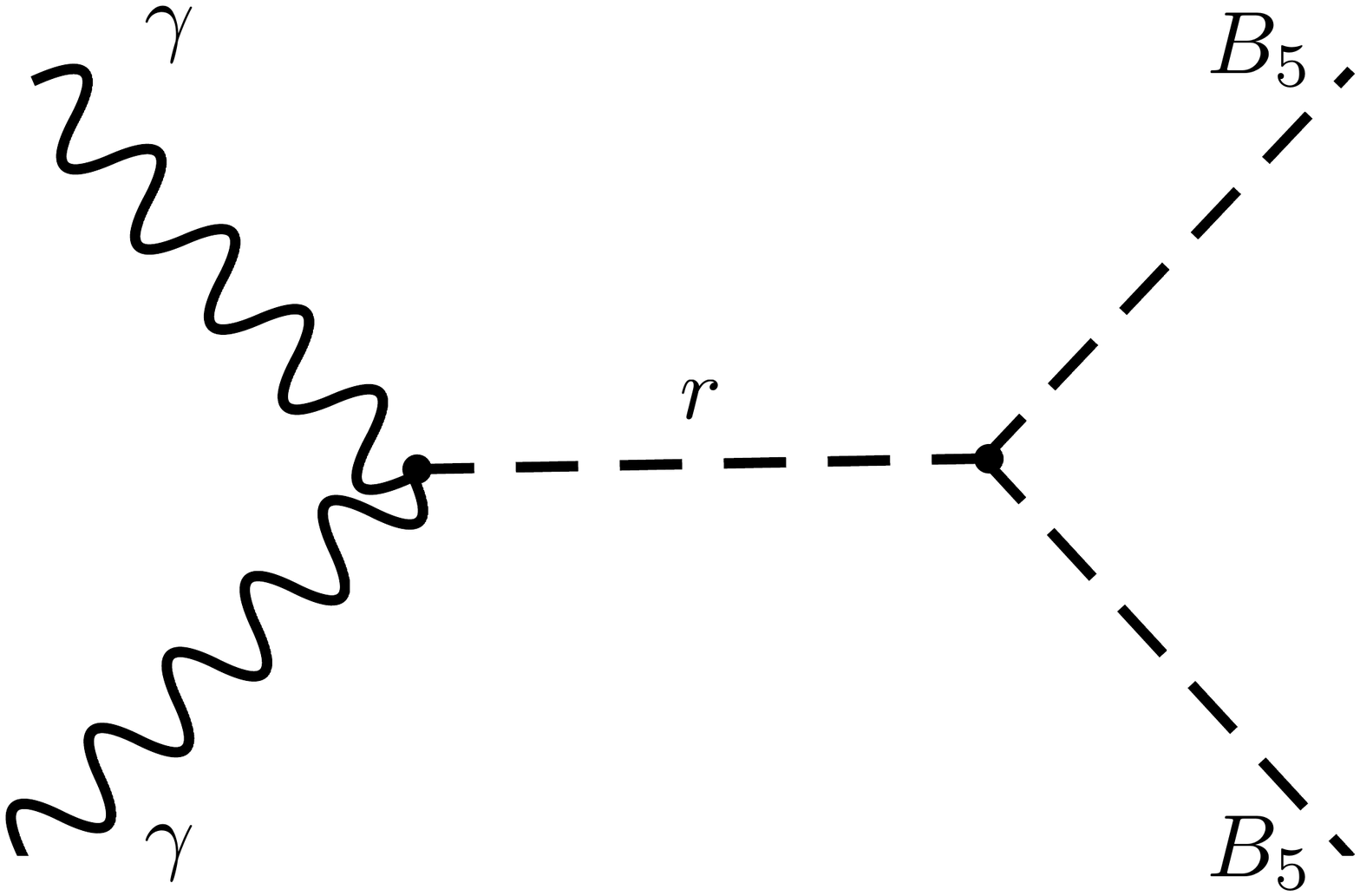} \includegraphics[width=2.5in]{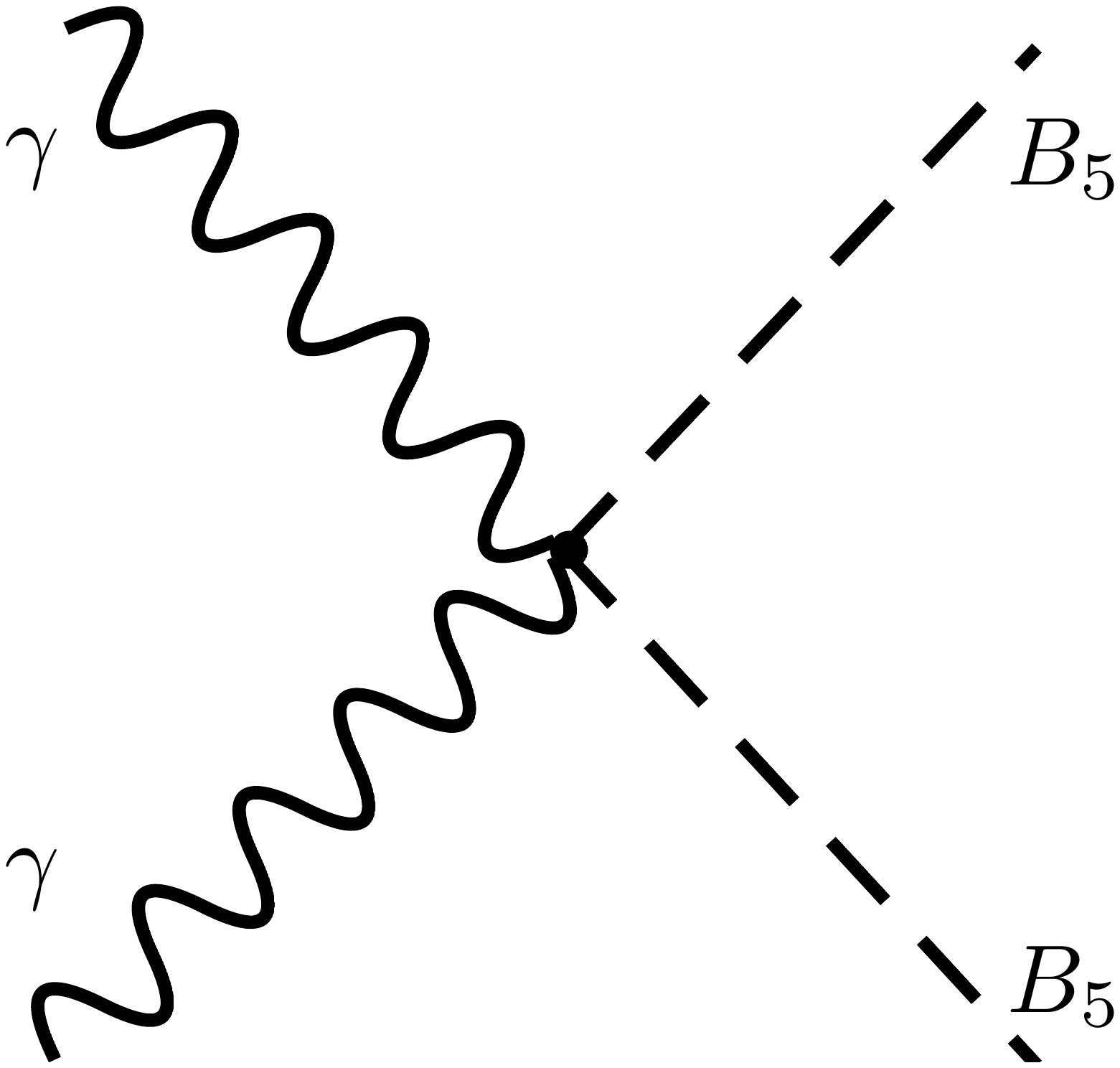}
   \caption{The diagram on the left involving the exchange of a RS radion leads to the effective dimension 8 contact operator shown on the right.}
\label{fig:ggb5b5}
\end{figure}

Diagrams such as the one shown in Figure~\ref{fig:ggb5b5} create higher dimensional operators in an effective theory valid at energies below the scale of RS gravitational excitations.  In this section, we calculate these higher dimensional operators as functions of the radion mass and the parameters associated with the RS geometry.

\begin{center} {\bf Operators for 5D Goldstone Bosons } \end{center}

The coefficients of the irrelevant operators arising from integrating out the radion can be determined by the form of the radion couplings to bulk SM fields~\cite{bulkradion}.  The results are given by:

\begin{eqnarray}
&&\mathcal{L}^{aa \gamma \gamma}_\text{eff} = \frac{ \left(\partial_\mu B_5 \right)^2 F_{\rho \sigma}^2}{4 m_r^2 \Lambda_r^2 \log R'/R} \nonumber \\
&&\mathcal{L}^{B_5 B_5 gg}_\text{eff} = \frac{ \left(\partial_\mu B_5 \right)^2 G_{\rho \sigma}^2}{4 m_r^2 \Lambda_r^2 \log R'/R} \nonumber \\
&&\mathcal{L}^{aa \bar{f} f}_\text{eff} = \frac{m_f (c_L - c_R)}{m_r^2 \Lambda_r^2}  \bar{f} f \left(\partial_\mu B_5 \right)^2
\end{eqnarray}
The last interaction is for fermions which are localized on the UV brane.  The coefficients $c_L$ and $c_R$ are the fermion bulk masses which determine the wave-functions of the zero modes.
The second interaction, at momentum transfer below the QCD scale, leads to an effective coupling of the Goldstone boson to nucleons:
\begin{equation}
\mathcal{L}^{B_5 B_5 n n}_\text{eff} = \frac{ \left(\partial_\mu B_5 \right)^2 \bar{n} n}{4 m_r^2 \Lambda_r^2 \log R'/R}  m_{n,p}  \frac{ 8 \pi }{ 9 \alpha_s } \left[ \sum_{q=u,d,s} f_{Tq} -1 \right]
\end{equation}
where $m_{n,p}$ is the neutron/proton mass.  The coefficient is obtained by taking the matrix element of the scalar gluon current between nucleons:
\begin{equation}
\bar{n} n \langle n | G_{\rho \sigma}^2 |n \rangle \rightarrow - \bar{n} n~m_n \frac{8 \pi}{9 \alpha_s} \left[ \sum_{q=u,d,s} f_{Tq} -1 \right]
\end{equation}
The $f_{Tq}$ coefficients are defined by $\langle n| m_q \bar{q} q | n \rangle \equiv m_n f_{Tq}$.


\begin{center} {\bf Operators for unbroken gauge symmetries} \end{center}

Similarly, there are higher dimensional operators involving massless bulk gauge fields, $B_\mu$.
\begin{eqnarray}
&&\mathcal{L}^{B B \gamma \gamma}_\text{eff} = \frac{ B_{\mu\nu}^2 F_{\rho \sigma}^2}{16 m_r^2 \Lambda_r^2 \log^2 R'/R} \nonumber \\
&&\mathcal{L}^{B B gg}_\text{eff} = \frac{ B_{\mu\nu}^2 G_{\rho \sigma}^2}{16 m_r^2 \Lambda_r^2 \log^2 R'/R} \nonumber \\
&&\mathcal{L}^{B B \bar{f} f}_\text{eff} = \frac{m_f (c_L - c_R)}{m_r^2 \Lambda_r^2}  \bar{f} f B_{\mu\nu}^2
\end{eqnarray}
These are invariant under the 4D gauge symmetry.  We leave a full analysis of the effects of these operators for future study.


\subsection{Main-Sequence Star and Supernova Cooling}

In massive astrophysical bodies, processes may occur which produce the light fields in within an RS hidden sector.  This is the case, for example, with standard axion scenarios, and which leads to significant constraints on the coupling strength of a pseudo-scalar axion to SM fields, $f_\text{PQ}^{-1}$.  However, our model predicts the existence of new TeV suppressed operators which can contribute to astrophysical pseudoscalar production.  If the HS fields are coupled weakly enough, the produced fields will free-stream out of the astrophysical body, and contribute in a straightforward way to its energy loss rate.  In main-sequence stars and supernovae, an increased energy-loss rate above that predicted within the SM has not been detected, putting constraints on the higher dimensional operators that arise from integrating out RS gravitational fluctuations.

In this section, we consider only RS hidden sectors containing a light Goldstone boson, not a light gauge field.  We leave constraints on HS gauge fields for future study.  These constraints are particularly relevant for the RS axion model considered in Section 5.

 We have calculated the scattering length of a 5D Goldstone boson produced in the core collapse neutron star phase of SN1987a, taking into account only the higher dimensional operators given above.   The scattering length is given approximately by
\begin{equation}
L = 1\cdot 10^{14} \text{m} \left( \frac{30~\text{MeV}}{E_a} \right)^{ 4} \left( \frac{ 1/R' } {500~\text{GeV}} \right)^4 \left( \frac{m_\text{radion}}{120~\text{GeV}} \right)^{4}
\end{equation}
This scattering length is far larger than the size of the core for reasonable choices of the parameters, and thus any produced Goldstone bosons in the core collapse process are free-streaming~\footnote{There are also other processes due to direct couplings of the Goldstone sector with the SM fields which can lead to re-scattering in a core collapse supernova.  However, the relevant range of allowed couplings for very light scalars $f_\text{eff}^-1$, are small enough to ensure that the light fields are still free streaming.}.

In a generic scattering process within a neutron star, where thermal conditions are semi-degenerate, the energy loss rate per unit volume due to particles which free-stream out of an object is given by
\begin{equation}
Q = \int d\Pi_\text{PS} S \left| \mathcal{M} \right|^2 \delta^{(4)} \left( \sum_i p_i - \sum_f p_f \right) E_\text{stream} f_1 f_2 \left(1- f_3 \right) \left( 1 - f_4 \right),
\end{equation}
where $E_\text{stream}$ is the energy lost in a single process due to particles streaming out of the object, and the $f_i$ are the thermal occupation functions of the neutrons and protons which scatter to produce the Goldstone bosons:
\begin{equation}
f_i = \frac{1}{e^{(E_i-\mu)/k T} + 1}
\end{equation}
The phase space integration is over both initial and final state particles, and $S$ contains initial and final state combinatorics for identical particles.

\begin{figure}[t]
   \centering
   \includegraphics[width=1.6in]{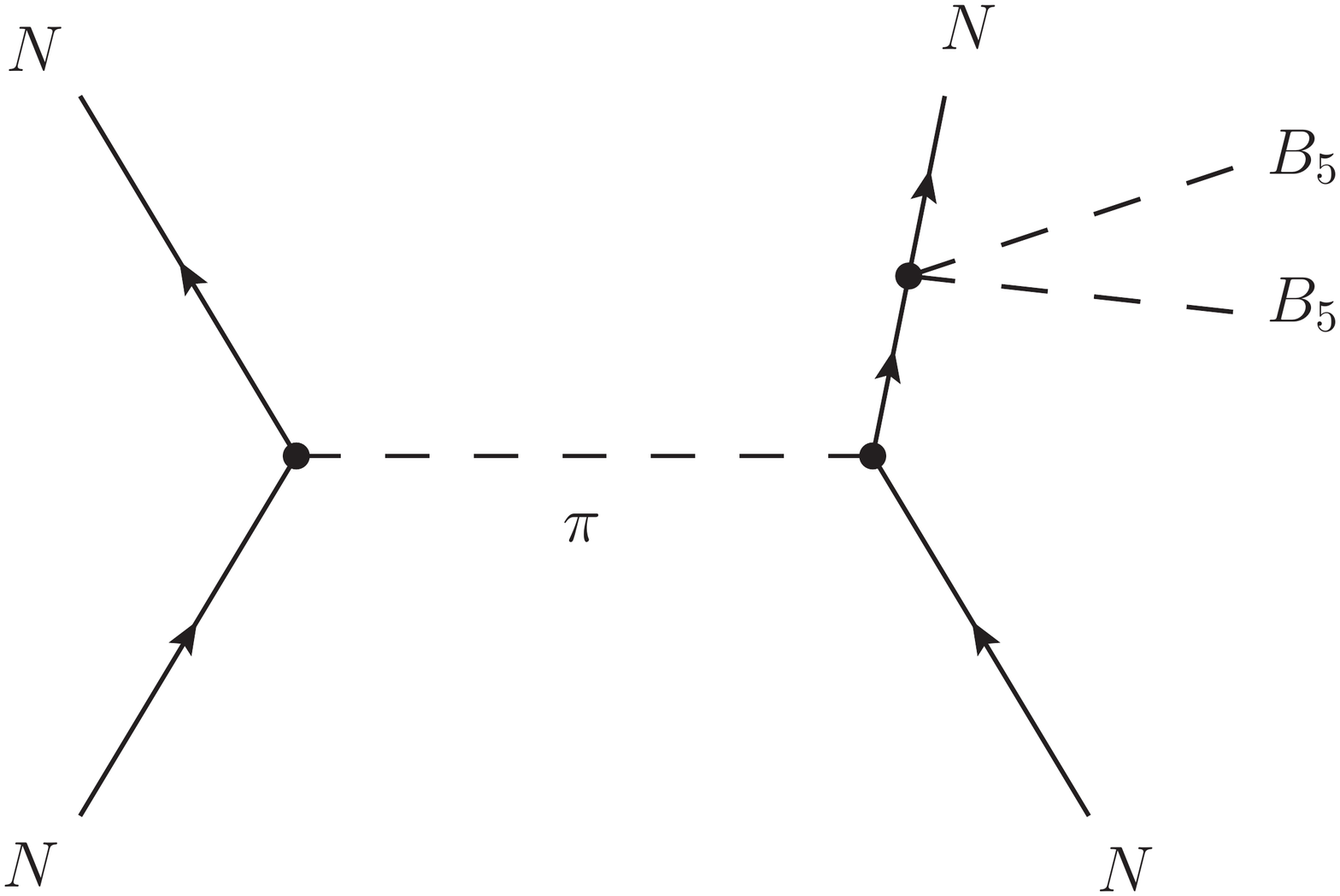}\includegraphics[width=1.6in]{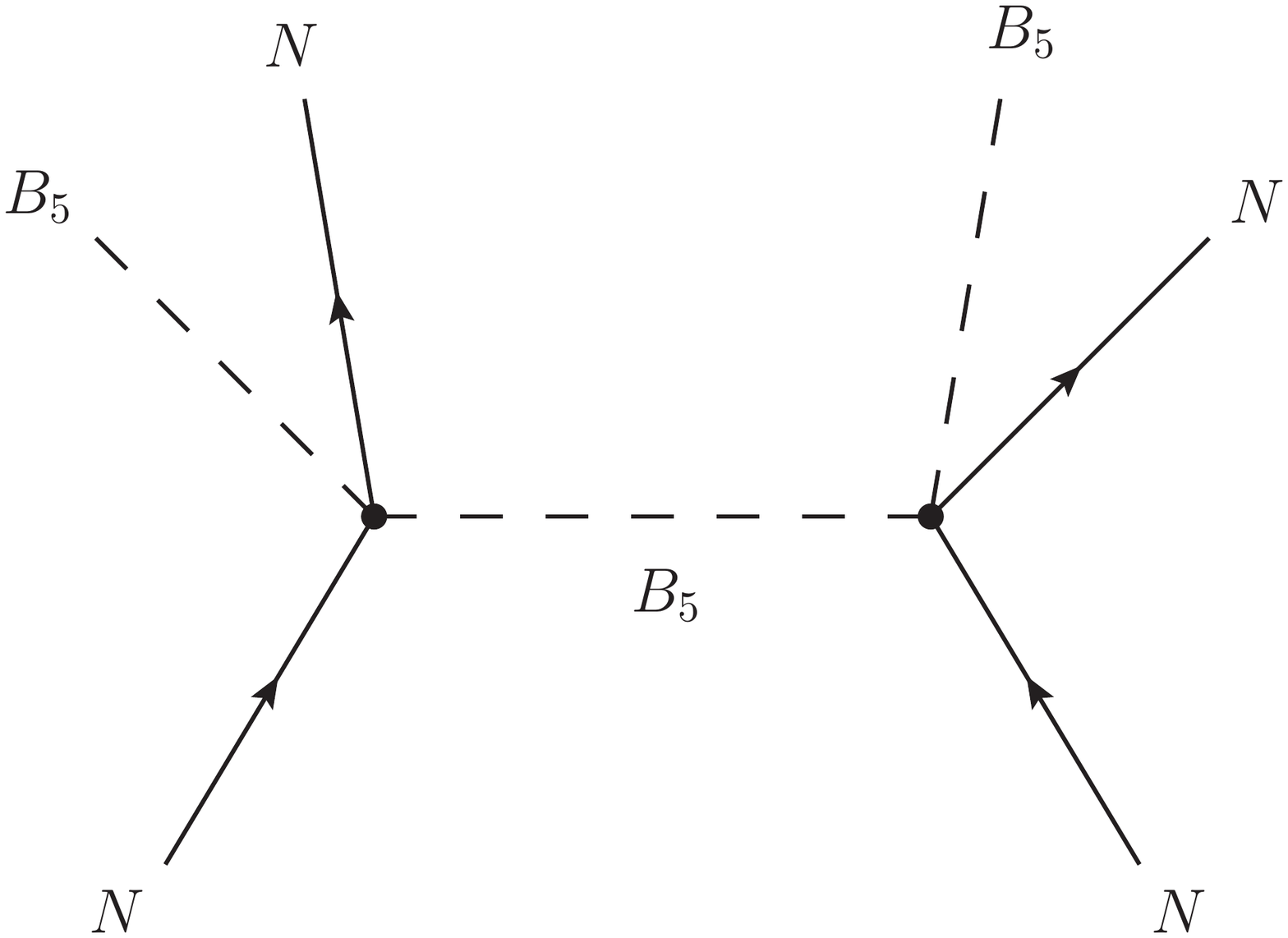}\includegraphics[width=1.6in]{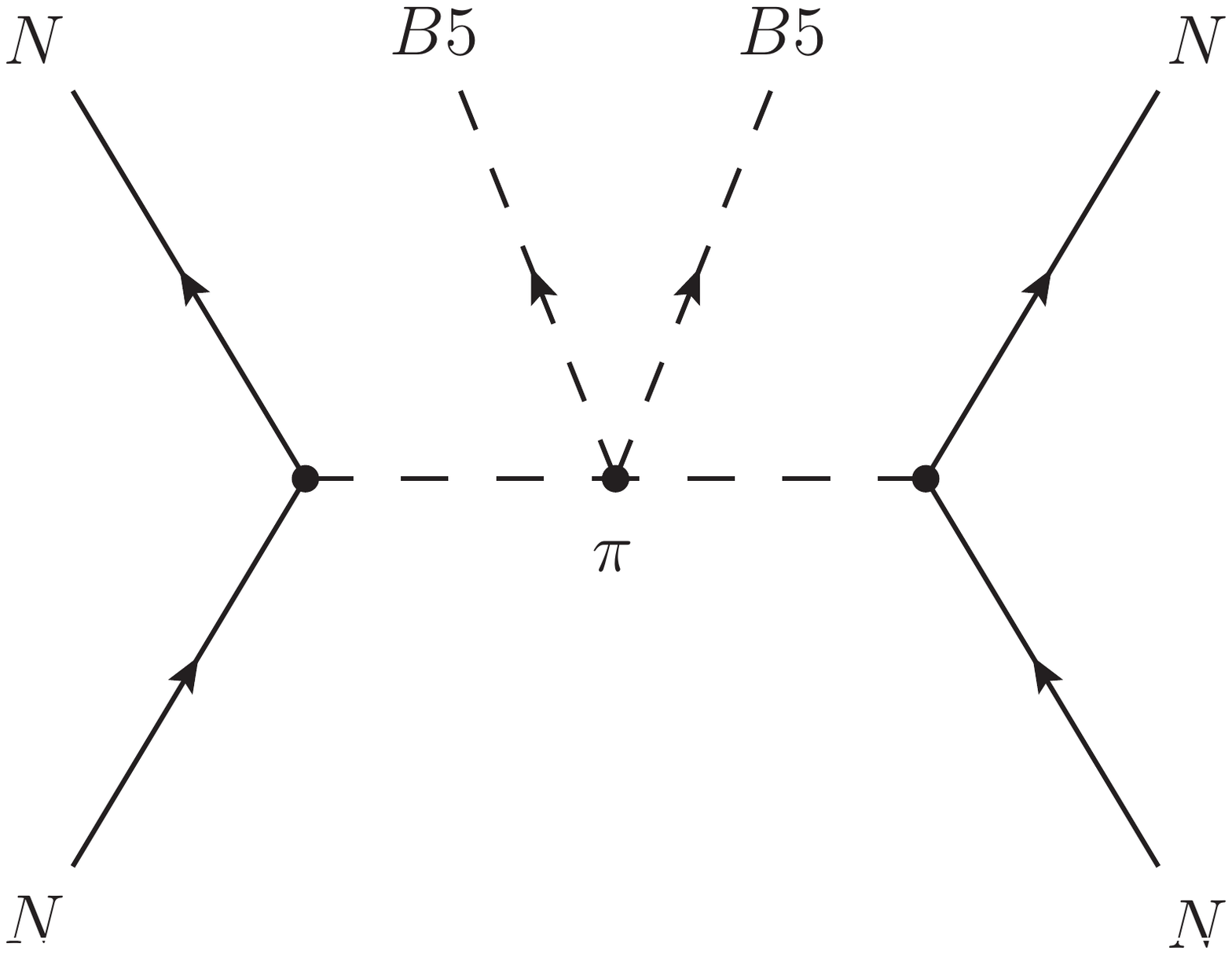}\includegraphics[width=1.6in]{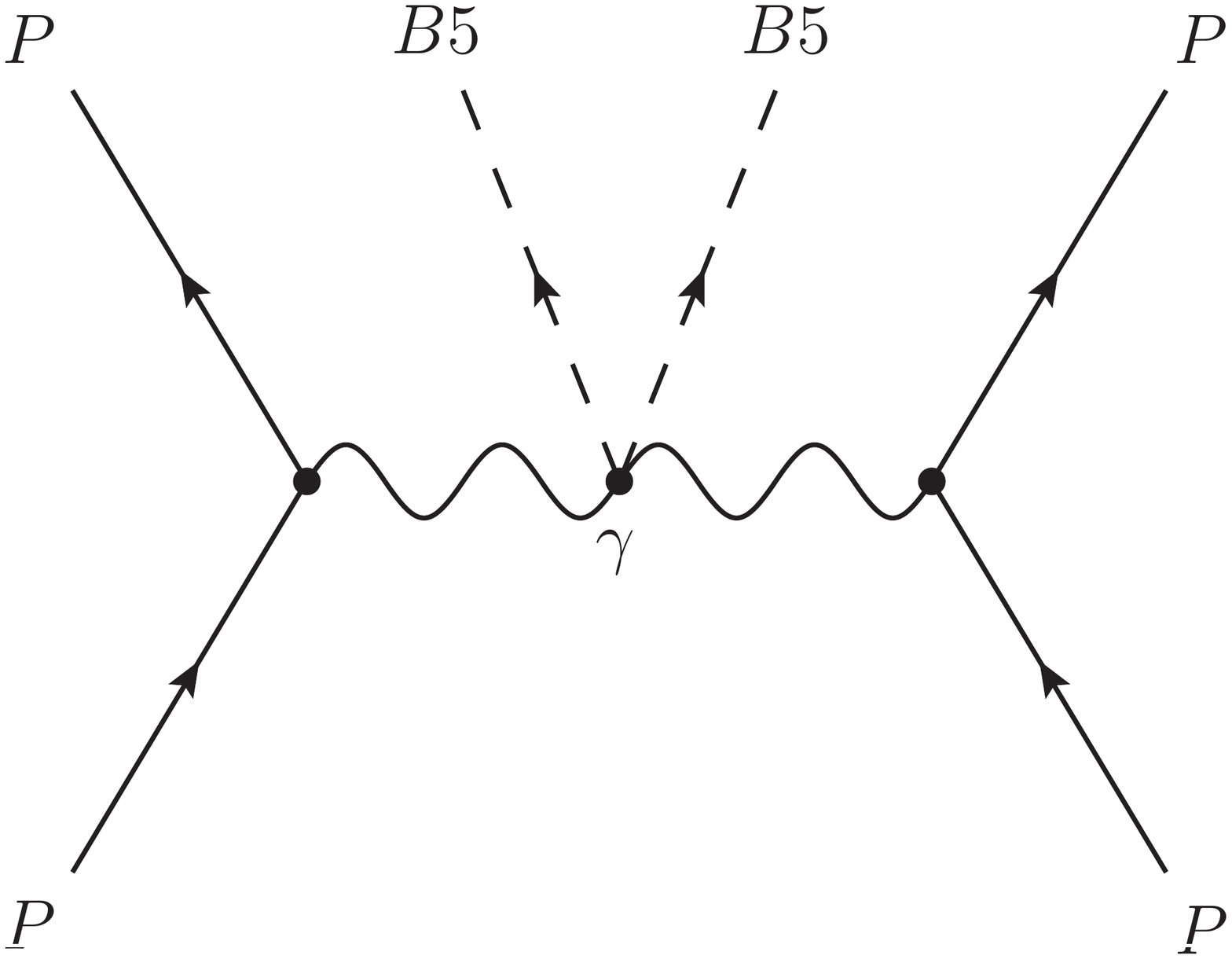}
   \caption{These are the new diagrams arising from RS gravitational excitations that contribute to supernova cooling.  $N$ is either a neutron or proton, while $P$ is a proton.  The higher dimensional operators involving $B_5$'s arise primarily from integrating out the radion.}
\label{fig:sncooling}
\end{figure}

We have estimated the energy loss rate due to nuclear bremsstrahlung in SN1987a.  The processes are: $n~n \rightarrow n~n~a~a$, where n is any nucleon, either a proton or neutron.  The diagrams which contribute to the matrix element are shown in Figure~\ref{fig:sncooling}.  We make a number of approximations in calculation the energy loss rate, but all of these simplifications overestimate the rate, meaning that the actual models are safer than what is reflected in our calculations.

First, we neglect the final state phase space distributions.  For Fermi-Dirac statistics, the $1-f_k$ functions vary between $1/2$ and $1$.  We take these to simply be one.  We overestimate the energy loss per collision by assuming it is equal to the total initial energy of the system:  $E_\text{stream} = E_\text{av} \equiv E_1+E_2-m_{n_1} - m_{n_2}$.  This means that the energy lost is purely a function of the initial states in the scattering process, and can be factored out of the final state phase space integration.  In reality, the energy lost is generally much less. Once this is done, the phase space integration over final states is the usual one for calculating cross sections:
\begin{eqnarray}
Q &=& \int d\Pi_\text{PS} S \left| \mathcal{M} \right|^2 \delta^{(4)} \left( \sum_i p_i - \sum_f p_f \right) E_\text{stream} f_1 f_2 \left(1- f_3 \right) \left( 1 - f_4 \right) \nonumber \\
&\lesssim& \int d\Pi_\text{is} E_\text{av} f_1 f_2 S \int d\Pi_\text{fs} \left| \mathcal{M} \right|^2 \delta^{(4)} \left( \sum_i p_i - \sum_f p_f \right) \nonumber \\
&=& \int d\Pi_\text{is} E_\text{av} f_1 f_2 S \left( 2 E_1 2 E_2 v_\text{rel} \sigma \right).
\end{eqnarray}
We compute the cross sections for the relevant processes using CalcHep~\cite{CalcHep}.  These are relativistically invariant functions of the center of mass energy of the collision, or equivalently the magnitudes of the 3-momenta in the center of mass frame.  We numerically interpolate these total cross sections over the relevant range of 3-momenta and perform the above integration numerically.

\begin{figure}[t]
   \centering
   \includegraphics[width=3in]{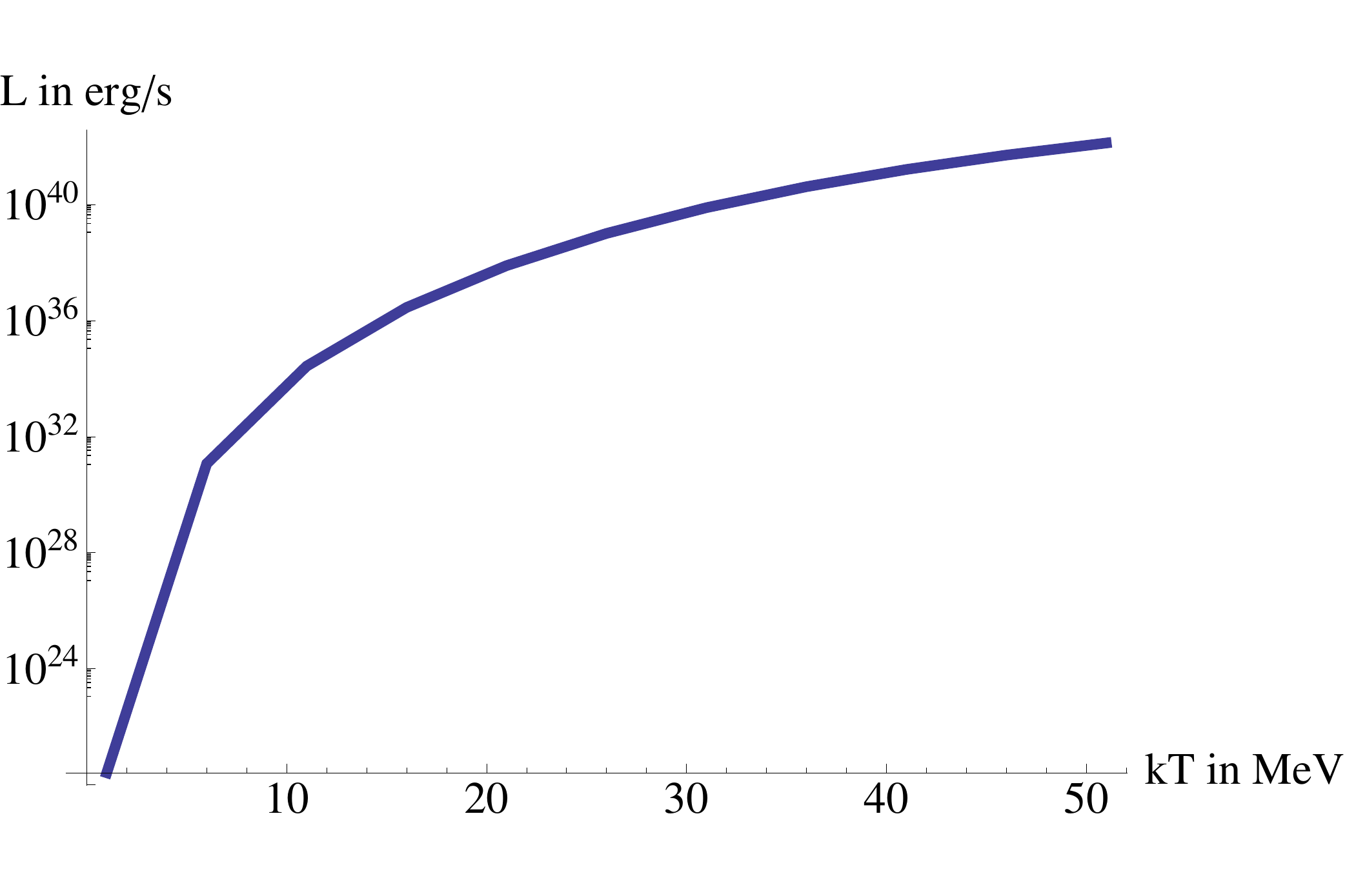}   \includegraphics[width=3in]{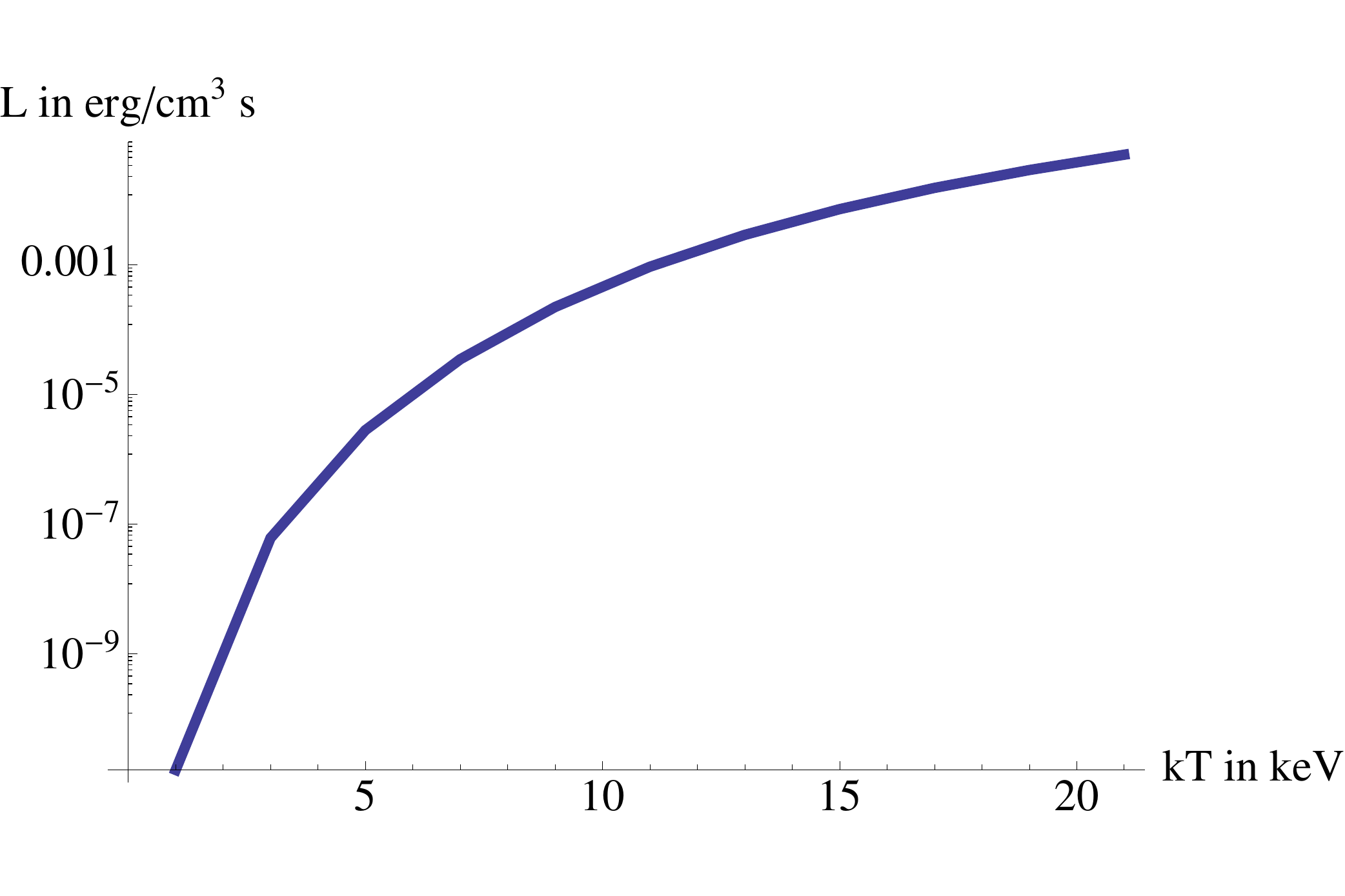}  \caption{In these plots, we show the temperature dependence of the luminosity in hidden sector goldstone bosons  during in a core collapse supernova (left) and in main sequence stars (right) for a single Goldstone scalar coupled to the SM via RS gravity excitations.  The following parameters are used:  $m_\text{radion} = 100~\text{GeV}$, $R' = (500 ~\text{GeV})^{-1}$, and $R = 1/M_\text{Pl}$.}
\label{fig:ktplot}
\end{figure}

The final energy loss rates due to the nuclear Bremsstrahlung processes are given by
\begin{eqnarray}
&&Q_\text{NN} = 3.9 \cdot 10^{20} \left( \frac{100~\text{GeV}}{m_\text{radion}} \right)^4 \left( \frac{36.8}{\log{R'/R}} \right)^2 \left( R'~500~\text{GeV} \right)^4 \text{erg}/\text{cm}^3/\text{s} \nonumber \\
&&Q_\text{PP} =  2.0 \cdot 10^{21}  \left( \frac{100~\text{GeV}}{m_\text{radion}} \right)^4 \left( \frac{36.8}{\log{R'/R}} \right)^2 \left( R'~500~\text{GeV} \right)^4\text{erg}/\text{cm}^3/\text{s} \nonumber \\
&&Q_\text{NP} = 3.9 \cdot 10^{20}  \left( \frac{100~\text{GeV}}{m_\text{radion}} \right)^4 \left( \frac{36.8}{\log{R'/R}} \right)^2 \left( R'~500~\text{GeV} \right)^4\text{erg}/\text{cm}^3/\text{s}
\end{eqnarray}
This corresponds to a total luminosity (for a $20$~km radius neutron star) of~$\mathcal{L}_a = 3 \cdot 10^{40} \text{erg}/\text{s}$, and temperature $k T = 30~\text{MeV}$ whereas the luminosity of the neutrino burst phase is estimated to be $\mathcal{L}_\nu \approx 10^{53} \text{erg}/\text{s}$.  Thus, for this choice of parameters, the additional energy loss due to processes involving the couplings of RS gravity to the HS can be neglected.   In Figure~\ref{fig:ktplot}, we display the temperature dependence of the total luminosity in Goldstone bosons due to the processes in Figure~\ref{fig:sncooling}.

\begin{figure}[t]
   \centering
   \includegraphics[width=1.6in]{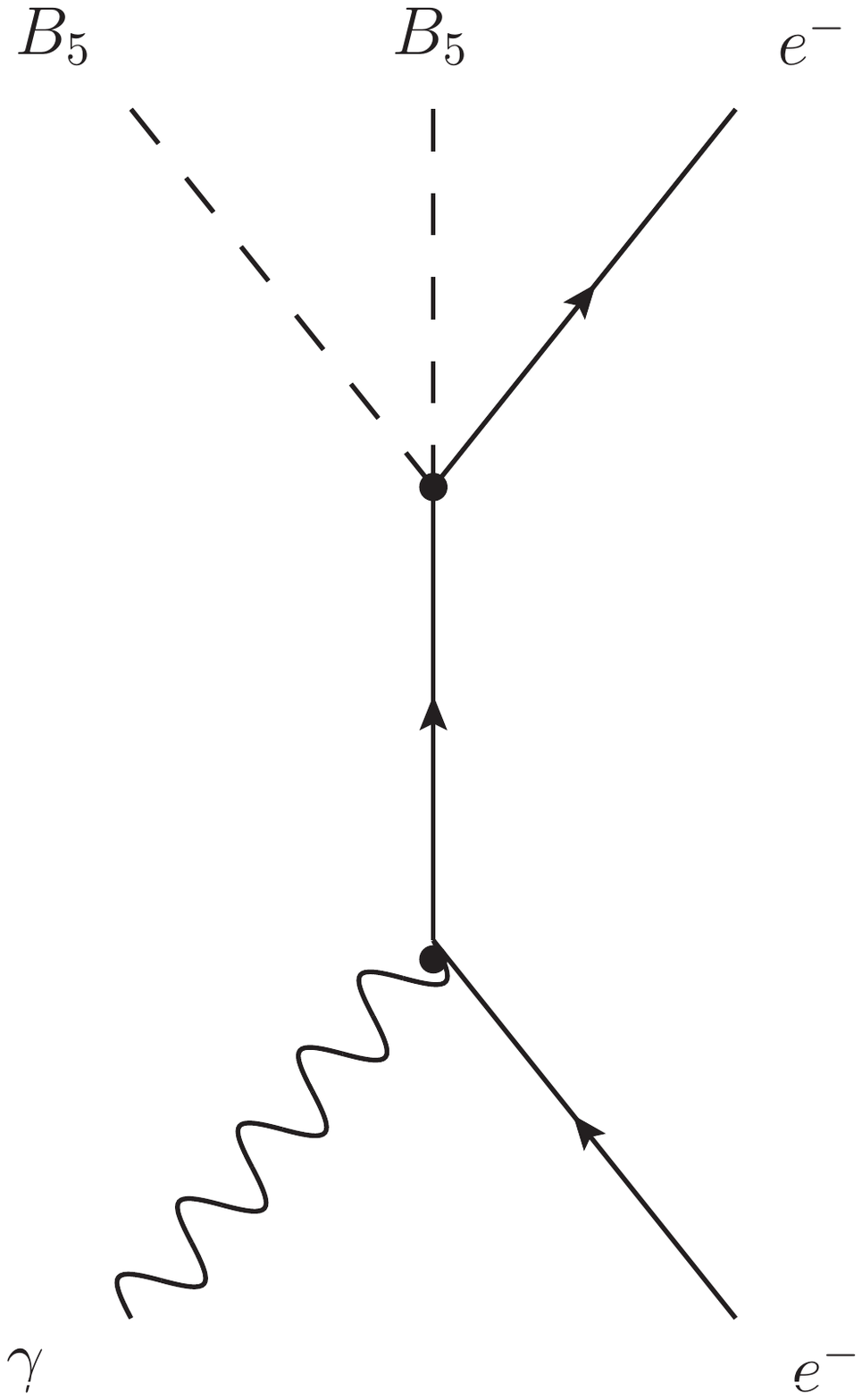}\includegraphics[width=1.6in]{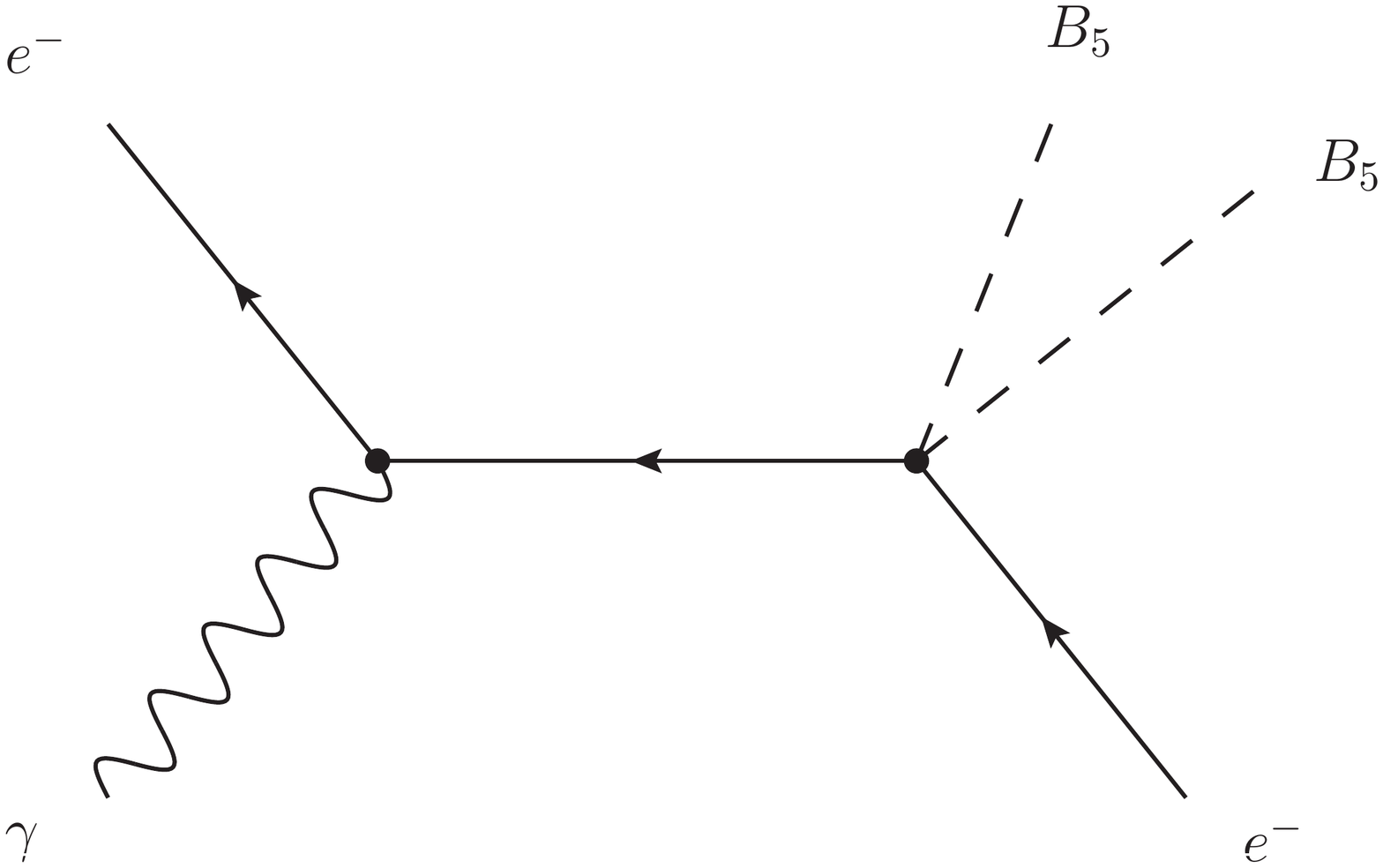}\includegraphics[width=1.6in]{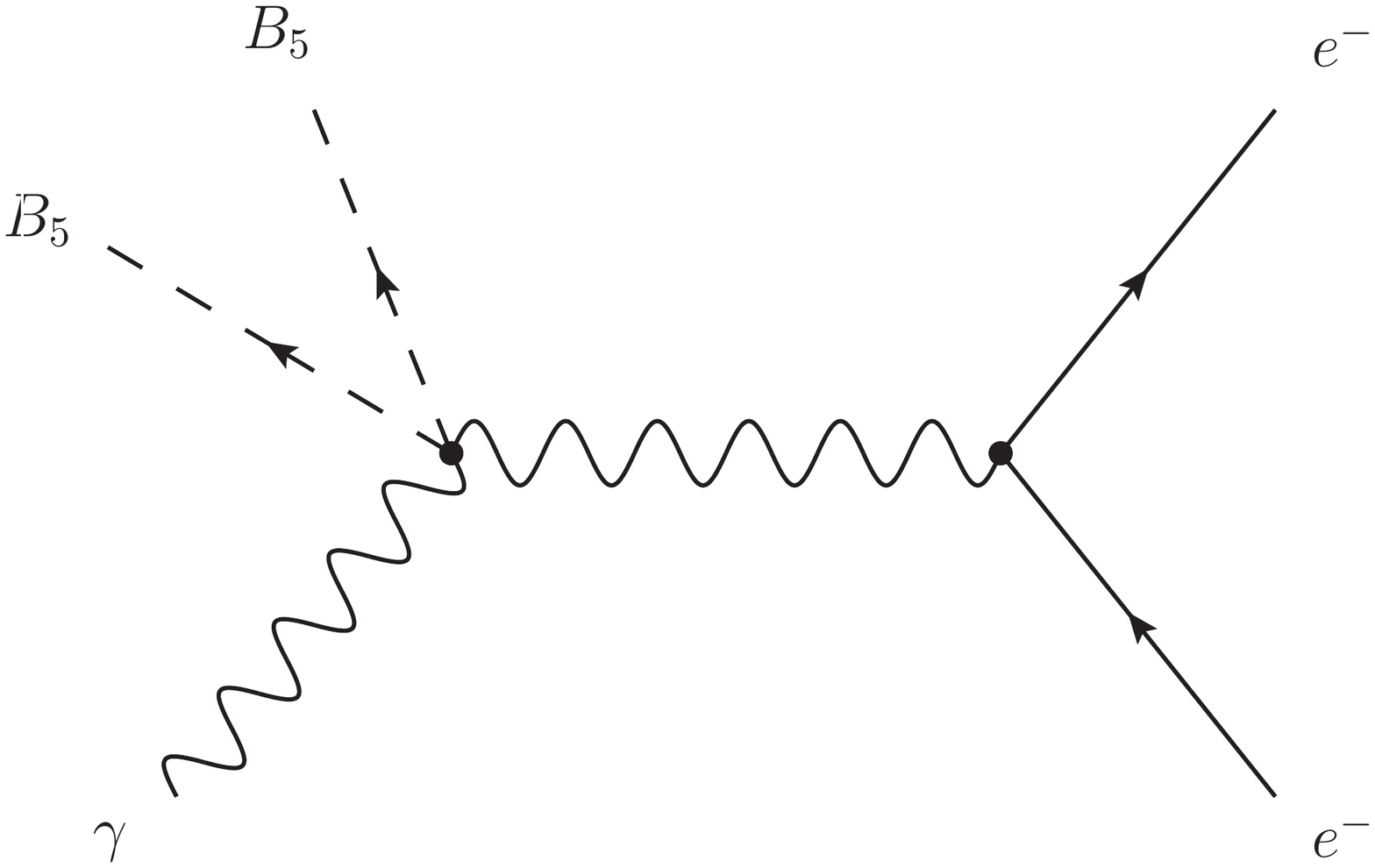}
   \caption{These are the new diagrams arising from RS gravitational excitations that contribute to star cooling.    In addition to these diagrams, the electron may be replaced by the nuclei of the solar elements.  The higher dimensional operators involving $B_5$'s arise primarily from integrating out the radion.}
\label{fig:starcooling}
\end{figure}

We have also calculated the energy loss rates in stars due to hidden Goldstone boson production from the processes shown in Figure~\ref{fig:starcooling}.  
Compton, Primakoff, and Bremsstrahlung diagrams contribute, as well as photon annihilation to Goldstone bosons.  The solar energy loss rates, using a temperature $k T = 1.3~\text{keV}$, for each process (labelled by the initial states) are given by:
\begin{eqnarray}
&&Q_{\gamma\gamma} = 6.7 \cdot 10^{-39}   \left( \frac{100~\text{GeV}}{m_\text{radion}} \right)^4 \left( \frac{36.8}{\log{R'/R}} \right)^2 \left( R'~500~\text{GeV} \right)^4 \text{erg}/\text{cm}^3/\text{s}  \nonumber \\
&&Q_{e^- \gamma} = 2.1 \cdot 10^{-36}  \left( \frac{100~\text{GeV}}{m_\text{radion}} \right)^4 \left( R'~500~\text{GeV} \right)^4\text{erg}/\text{cm}^3/\text{s} \nonumber
\end{eqnarray}
\begin{eqnarray}
&&Q_{H \gamma} =  6.8 \cdot 10^{-11}  \left( \frac{100~\text{GeV}}{m_\text{radion}} \right)^4 \left( \frac{36.8}{\log{R'/R}} \right)^2 \left( R'~500~\text{GeV} \right)^4\text{erg}/\text{cm}^3/\text{s} \nonumber \\
&&Q_{He \gamma} = 1.1 \cdot 10^{-10}  \left( \frac{100~\text{GeV}}{m_\text{radion}} \right)^4 \left( \frac{36.8}{\log{R'/R}} \right)^2 \left( R'~500~\text{GeV} \right)^4 \text{erg}/\text{cm}^3/\text{s}
\end{eqnarray}
The Compton process has a different scaling due to the fact that the $B_5$ couplings to electrons are not dependent on the $\log$ of the scale hierarchy.  In comparison with usual solar nuclear energy production of a few $\text{erg}/\text{cm}^3/\text{s}$, these energy loss rates are negligible.  In Figure~\ref{fig:ktplot} we display the temperature dependence of the total energy loss rate, so that the results can be extended to other main-sequence stars.  For the higher red-giant core temperatures, the energy loss rate is still small in comparison with nuclear burning rates of about $10^{8}~\text{erg}/\text{cm}^3 \text{s}$.


\section*{Conclusions}

We have examined a class of models embedded in a Randall-Sundrum geometry in which there are new extra dimensional gauge symmetries which contain in their spectra either light scalar fields or light gauge fields.  These new fields are taken to be hidden from the SM, either through small couplings, or vanishing quantum numbers.  Such hidden sectors are still phenomenologically relevant, however, due to sizable couplings to RS gravitational fluctuations which, in turn, couple with similar strength to SM fields.  Through these couplings, the collider phenomenology of the radion and KK-gravitons may be drastically modified, and through scalar mixing, Higgs phenomenology may change as well.  We also motivate the case for such a hidden sector by describing a simple model which resolves the strong CP problem, and in which a light scalar field arising from an RS gauge symmetry plays the role of an axion.  Hidden sectors which contain such light scalar fields contribute new amplitudes relevant for star and supernova cooling.   We have calculated constraints arising from these operators, and find them to be well within current bounds.

\section*{Acknowledgements}

The authors thank Csaba Cs\'aki and Johannes Heinonen for discussion throughout the preparation of this manuscript.  JH thanks Matthew Strassler and Liam McAllister for discussions relevant to limits on the size of extra dimensional gauge couplings.  JH and DB thank the Institute of High Energy Phenomenology at Cornell University for generous hospitality throughout this work.  JH thanks the Aspen Center for Physics and DB thanks the TASI summer school for hospitality as well.  The work of JH and DB has been supported by the College of Arts and Sciences at Syracuse University.

\section{Appendix A:  Tables of gravitational interactions}
\setcounter{table}{0}
In this Appendix, we summarize the interactions of the radion and the gravitational excitations with both broken and unbroken 5D gauge symmetries. 

 In Table~\ref{tab:gravcoupGS}, we give the interactions of the radion and graviton KK-modes with the massless $B_5$ and the associated KK-modes in the case where the gauge symmetry is broken twice by boundary conditions.  In Table~\ref{tab:gravcoupRUB}, we give the couplings of the radion to an unbroken gauge group.  Finally, in Table~\ref{tab:gravcouphUB}, we give the couplings of the first two KK-gravitons to an unbroken bulk gauge group.

\begin{table}[h]
\center{\begin{tabular}{|l|c||l|c||l|c|}
\hline
$r B^{(1)\mu} \partial_\mu B_5                                             $       &  $  1.09 \frac{M_1}{\kappa \Lambda_r} $&
$\hat{h}_{(1)}^{\mu\nu} B^{(1)}_\mu \partial_\nu B_5       $      &  $ -0.134 \frac{M_1}{\kappa \Lambda_1}$  &
$\hat{h}_{(2)}^{\mu\nu} B^{(1)}_\mu \partial_\nu B_5       $      &  $ .099 \frac{M_1}\kappa {\Lambda_2} $ \\ \hline
$r B^{(1)}_\mu B^{(1)\mu}                                                      $      &  $  \frac{4}{3} \frac{M_1^2}{2 \kappa \Lambda_r}$&
$\hat{h}_{(1)}^{\mu\nu} B^{(1)}_\mu B^{(1)}_\nu                $     &  $  -.137 \frac{M_1^2}{2 \kappa \Lambda_1}$&
$\hat{h}_{(2)}^{\mu\nu} B^{(1)}_\mu B^{(1)}_\nu                $     &  $  .050 \frac{M_1^2}{2 \kappa \Lambda_2}$\\\hline
$r B^{(1)}_{\mu\rho} B^{(1)\mu\rho}                                       $    &  $ \frac{1}{3} \frac{1}{2 \kappa \Lambda_r}$&
$\hat{h}_{(1)}^{\mu\nu} B^{(1)}_{\mu\rho} B^{(1)\rho}_{\nu}$  &  $  .137 \frac{1}{2 \kappa \Lambda_1}$&
$\hat{h}_{(2)}^{\mu\nu} B^{(1)}_{\mu\rho} B^{(1)\rho}_{\nu} $ &  $  .053 \frac{1}{2 \kappa \Lambda_2}$\\\hline
$r ( \partial_\mu B_5)^2                                                               $ &  $  2 \frac{1}{2 \kappa \Lambda_r}$&
$\hat{h}_{(1)}^{\mu\nu}  \partial_\mu B_5 \partial_\nu B_5  $  &  $  -.219 \frac{1}{2 \kappa \Lambda_1}$&
$\hat{h}_{(2)}^{\mu\nu}  \partial_\mu B_5 \partial_\nu B_5  $  &  $  .049 \frac{1}{2 \kappa \Lambda_2}$ \\
\hline
\end{tabular}}
\caption{This table contains the Lagrangian coefficients for interactions between RS gravitational excitations and the modes associated with the bulk gauge symmetry that produces light Goldstone modes.}
\label{tab:gravcoupGS}
\end{table}

\begin{table}[h]
\center{\begin{tabular}{|l|c||l|c||l|c|}
\hline
$r B^{(0)}_{\mu\nu} B^{(0)\mu\nu}$               &    $\frac{1}{4 \kappa \Lambda_r \log R'/R}$ &
$r B^{(0)}_{\mu\nu} B^{(1)\mu\nu}$               &   $.483 \frac{1}{ \kappa \Lambda_r \sqrt{ \log R'/R}}$ &
$r B^{(0)}_{\mu\nu} B^{(2)\mu\nu}$               &   $-.090 \frac{1}{ \kappa \Lambda_r  \sqrt{\log R'/R}}$
\\ \hline 
$r B^{(1)}_{\mu\nu} B^{(1)\mu\nu}$               &    $.556 \frac{1}{2 \kappa \Lambda_r}$ &
$r B^{(1)}_{\mu} B^{(1)\mu}$                           &   $.222 \frac{M_1^2}{2 \kappa \Lambda_r}$ &
$r B^{(1)}_{\mu\nu} B^{(2)\mu\nu}$               & $-.237 \frac{1}{ \kappa \Lambda_r }$
\\ \hline 
$r B^{(1)}_\mu B^{(2)\mu}$                             & $-.175 \frac{M_1 M_2}{ \kappa \Lambda_r}$ &
$r B^{(2)}_{\mu\nu} B^{(2)\mu\nu}$                & $.377 \frac{1}{2 \kappa \Lambda_r}$ &
$r B^{(2)}_{\mu} B^{(2)\mu}$                           &   $.312 \frac{M_2^2}{2 \kappa \Lambda_r}$ 
\\ \hline
\end{tabular}}
\caption{This table contains the Lagrangian coefficients for interactions between the radion and the zero and KK-modes of an unbroken RS gauge symmetry.}
\label{tab:gravcoupRUB}
\end{table}

\begin{table}[h]
\center{\begin{tabular}{|l|c||l|c||l|c|}
\hline
$h_{(1)}^{\mu\nu} B^{(0)}_{\mu\rho} B^{(0)\rho}_\nu$               &    $.191 \frac{1}{2 \kappa \Lambda_1 \log R'/R}$ &
$h_{(1)}^{\mu\nu} B^{(0)}_{\mu\rho} B^{(1)\rho}_\nu$               &   $.209 \frac{1}{ \kappa \Lambda_1 \sqrt{\log R'/R}}$ &
$h_{(1)}^{\mu\nu} B^{(0)}_{\mu\rho} B^{(2)\rho}_\nu$               & $-.066 \frac{1}{ \kappa \Lambda_1  \sqrt{\log R'/R}}$
\\ \hline 
$h_{(1)}^{\mu\nu} B^{(1)}_{\mu\rho} B^{(1)\rho}_\nu$                &    $.249 \frac{1}{2 \kappa \Lambda_1}$ &
$h_{(1)}^{\mu\nu} B^{(1)}_{\mu} B^{(1)}_\nu$                           &   $-.082 \frac{M_1^2}{2  \kappa \Lambda_1}$ &
$h_{(1)}^{\mu\nu} B^{(1)}_{\mu\rho} B^{(2)\rho}_\nu$               & $-.118 \frac{1}{ \kappa \Lambda_1 }$
\\ \hline 
$h_{(1)}^{\mu\nu} B^{(1)}_\mu B^{(2)}_\nu$                             & $.086 \frac{M_1 M_2}{ \kappa \Lambda_1}$ &
$h_{(1)}^{\mu\nu} B^{(2)}_{\mu\rho} B^{(2)\rho}_\nu$                & $.152 \frac{1}{2 \kappa \Lambda_1}$ &
$h_{(1)}^{\mu\nu} B^{(2)}_{\mu} B^{(2)}_\nu$                           &   $-.138 \frac{M_2^2}{2 \kappa \Lambda_1}$ 
\\ \hline
$h_{(2)}^{\mu\nu} B^{(0)}_{\mu\rho} B^{(0)\rho}_\nu$               &    $.028 \frac{1}{2 \kappa \Lambda_2 \log R'/R}$ &
$h_{(2)}^{\mu\nu} B^{(0)}_{\mu\rho} B^{(1)\rho}_\nu$               &   $-.024 \frac{1}{ \kappa \Lambda_2 \sqrt{\log R'/R}}$ &
$h_{(2)}^{\mu\nu} B^{(0)}_{\mu\rho} B^{(2)\rho}_\nu$               & $.119 \frac{1}{ \kappa \Lambda_2  \sqrt{\log R'/R}}$
\\ \hline 
$h_{(2)}^{\mu\nu} B^{(1)}_{\mu\rho} B^{(1)\rho}_\nu$                &    $-.064 \frac{1}{2 \kappa \Lambda_2}$ &
$h_{(2)}^{\mu\nu} B^{(1)}_{\mu} B^{(1)}_\nu$                           &   $-.051 \frac{M_1^2}{2 \kappa \Lambda_2}$ &
$h_{(2)}^{\mu\nu} B^{(1)}_{\mu\rho} B^{(2)\rho}_\nu$               & $.116 \frac{1}{ \kappa \Lambda_2 }$
\\ \hline 
$h_{(2)}^{\mu\nu} B^{(1)}_\mu B^{(2)}_\nu$                             & $.0031 \frac{M_1 M_2}{ \kappa \Lambda_2}$ &
$h_{(2)}^{\mu\nu} B^{(2)}_{\mu\rho} B^{(2)\rho}_\nu$                & $-.282 \frac{1}{2 \kappa \Lambda_2}$ &
$h_{(2)}^{\mu\nu} B^{(2)}_{\mu} B^{(2)}_\nu$                           &   $-5.9\cdot 10^{-4} \frac{M_2^2}{2  \kappa \Lambda_2}$ 
\\ \hline
\end{tabular}}
\caption{This table contains the Lagrangian coefficients for interactions between the KK-gravitons and the zero and KK-modes of an unbroken RS gauge symmetry.}
\label{tab:gravcouphUB}
\end{table}

\section*{Appendix B:  Gauge fixing of the Hidden Sector}
Since we are including the coupling of gravity to the gauge fields, and we have already chosen a specific gauge in which to express the gravitational fluctuations, we must be sure to respect general covariance in the gauge fixing term we add to restrict the path integral to non-redundant hidden sector gauge field configurations.  This is to ensure we do not create spurious interactions which are artifacts of over-constraining the gauge freedom.  Note that the general R-$\xi$ gauges often chosen in such models break 5D covariance, even in the bulk, so we must find a new gauge fixing potential.  The one we choose is, in the end, equivalent at the quadratic level to the 5D R-$\xi$ gauges~\cite{gaugefixing} with the choice $\xi = 1$, however the non-covariant R-$\xi$ gauge still generates spurious 3-point couplings involving KK-gravitons and the radion.

To begin, we write the gauge kinetic term in an explicitly covariant manner (although as usual the Christoffel symbols cancel by anti-symmetry of the gauge field strength tensor):
\begin{eqnarray}
S_\text{gauge} &=& -\frac{1}{4 g_5^2} \int_\mathcal{M} dV g^{MN} g^{RS} \left( \nabla_M A_R -\nabla_R A_M \right)  \left( \nabla_N A_S -\nabla_S A_N \right) \nonumber \\
 &=& \frac{1}{2 g_5^2} \int_\mathcal{M} dV g^{MN} g^{RS} \left( \nabla_R A_M \nabla_N A_S - \nabla_M A_R \nabla_N A_S \right)
\end{eqnarray}
where $dV$ is the covariant volume element.   We would ideally like to remove the kinetic mixing between the vector fields and the components which are eaten to produce massive 4D vectors in the effective field theory.

A general covariant gauge fixing term which removes the mixing is given by:
\begin{equation}
\mathcal{S}_{GF} = -  \frac{1}{2 g_5^2} \int_\mathcal{M} dV \mathcal{G}(B)^2 =  -  \frac{1}{2 g_5^2} \int_\mathcal{M} dV \left( \nabla_M A^M + v_M A^M \right)^2
\end{equation}
Here, $v_M$ is a vector field whose components we will determine in this section.  
Expanded in component form, in the absence of gravity fluctuations, this gauge fixing function is:
\begin{equation}
\left(\frac{R}{z}\right)^2 \left[ \partial_\mu B_\nu \eta^{\mu\nu} - B'_5 + 3 \frac{B_5}{z} +\eta^{\mu\nu} v_\mu B_\nu - v_5 B_5 \right]
\end{equation}

The residual gauge symmetry with this gauge fixing term obeys the following equation (in the absence of gravity fluctuations):
\begin{equation}
\Box \beta - \beta'' + 3 \frac{\beta'}{z} + \eta^{\mu \nu} v_\mu \partial_\nu \beta - v_5 \beta' =0
\end{equation}

The kinetic mixing term between $B_\mu$ and $B_5$, after summing up the standard kinetic term and the contributions from the gauge fixing term are:
\begin{equation}
\frac{1}{g_\text{5D}^2}\left( \frac{R}{z} \right) \left[ \left( \partial_\mu B_\nu \eta^{\mu\nu}  +\eta^{\mu\nu} v_\mu B_\nu \right) \left( B'_5 - 3 \frac{B_5}{z}+ v_5 B_5 \right) - B'_\mu \partial_\nu B_5 \eta^{\mu\nu} \right]
\end{equation}
Integration by parts of the last term in this expression causes the entire mixing term to vanish if the vector $v_M$ is chosen such that $v_\mu = 0$, and $v_5 = 2/z$.

Note that the gauge fixing function $\mathcal{G}(B)$ is a function of $\partial_\mu B^\mu$ and only the eaten $B_5$ modes.  The variation of the gauge fixing term then, with respect to the metric, is:
\begin{equation}
\frac{\delta}{\delta g^{MN}} \mathcal{L}_\text{GF} = -\frac{1}{2 g_\text{5D}^2} \left( \frac{\delta}{\delta g^{MN}} \sqrt{g} \right) \mathcal{G}(B)^2 + \sqrt{g} \left( \frac{\delta}{\delta g^{MN}} \mathcal{G}(B) \right) \mathcal{G}(B)
\end{equation}
Thus all interactions with gravitational fluctuations involve only the unphysical $B_5$'s, and terms involving $\partial_\mu B^\mu$ which vanish in all matrix elements due to the 4D Ward identities for the HS KK-modes.  The interactions listed in Appendix A are thus sufficient to describe the physical couplings of RS gravity to the excitations within the HS.



\begin{thebibliography}{99}

\bibitem{RS}
  L.~Randall and R.~Sundrum,
  Phys.\ Rev.\ Lett.\  {\bf 83}, 3370 (1999)
  [arXiv:hep-ph/9905221].

\bibitem{HL1}
  C.~Csaki, C.~Grojean, H.~Murayama, L.~Pilo and J.~Terning,
  Phys.\ Rev.\  D {\bf 69}, 055006 (2004)
  [arXiv:hep-ph/0305237].
  C.~Csaki, C.~Grojean, L.~Pilo and J.~Terning,
  Phys.\ Rev.\ Lett.\  {\bf 92}, 101802 (2004)
  [arXiv:hep-ph/0308038].
  C.~Csaki and D.~Curtin,
  Phys.\ Rev.\  D {\bf 80}, 015027 (2009)
  [arXiv:0904.2137 [hep-ph]].

\bibitem{HL2}
  C.~Csaki, C.~Grojean, J.~Hubisz, Y.~Shirman and J.~Terning,
  Phys.\ Rev.\  D {\bf 70}, 015012 (2004)
  [arXiv:hep-ph/0310355].

\bibitem{HL3}
  C.~Csaki, J.~Hubisz and P.~Meade,
  arXiv:hep-ph/0510275.

\bibitem{CH1}
  R.~Contino, Y.~Nomura and A.~Pomarol,
  Nucl.\ Phys.\  B {\bf 671}, 148 (2003)
  [arXiv:hep-ph/0306259].

\bibitem{CH2}
  K.~Agashe, R.~Contino and A.~Pomarol,
  Nucl.\ Phys.\  B {\bf 719}, 165 (2005)
  [arXiv:hep-ph/0412089].

\bibitem{strongCP}
  G.~'t Hooft,
  Phys.\ Rev.\ Lett.\  {\bf 37}, 8 (1976).

\bibitem{orbifoldBC}
  Y.~Kawamura,
  Prog.\ Theor.\ Phys.\  {\bf 105}, 999 (2001)
  [arXiv:hep-ph/0012125].

\bibitem{TASI}
  R.~Sundrum,
  arXiv:hep-th/0508134.
  C.~Csaki,
  arXiv:hep-ph/0404096.
  C.~Csaki, J.~Hubisz and P.~Meade,
  arXiv:hep-ph/0510275.


\bibitem{graviscalars}
  G.~F.~Giudice, R.~Rattazzi and J.~D.~Wells,
  Nucl.\ Phys.\  B {\bf 595}, 250 (2001)
  [arXiv:hep-ph/0002178].
  
\bibitem{radcsaba}
  C.~Csaki, M.~L.~Graesser and G.~D.~Kribs,
  Phys.\ Rev.\  D {\bf 63}, 065002 (2001)
  [arXiv:hep-th/0008151].

\bibitem{kkgravphen}
  A.~L.~Fitzpatrick, J.~Kaplan, L.~Randall and L.~T.~Wang,
  JHEP {\bf 0709}, 013 (2007)
  [arXiv:hep-ph/0701150].
  K.~Agashe, H.~Davoudiasl, G.~Perez and A.~Soni,
  Phys.\ Rev.\  D {\bf 76}, 036006 (2007)
  [arXiv:hep-ph/0701186].
  
\bibitem{bulkradion}
  C.~Csaki, J.~Hubisz and S.~J.~Lee,
  Phys.\ Rev.\  D {\bf 76}, 125015 (2007)
  [arXiv:0705.3844 [hep-ph]].

\bibitem{HV1}
  M.~J.~Strassler and K.~M.~Zurek,
  Phys.\ Lett.\  B {\bf 651}, 374 (2007)
  [arXiv:hep-ph/0604261].
  
\bibitem{HV2}
  K.~M.~Zurek,
  arXiv:1001.2563 [hep-ph].

\bibitem{PQ}
  R.~D.~Peccei and H.~R.~Quinn,
  Phys.\ Rev.\ Lett.\  {\bf 38}, 1440 (1977).
  

\bibitem{WWaxion}
  S.~Weinberg,
  Phys.\ Rev.\ Lett.\  {\bf 40}, 223 (1978).
  F.~Wilczek,
  Phys.\ Rev.\ Lett.\  {\bf 40}, 279 (1978).

\bibitem{warpedaxions}
  K.~w.~Choi,
  Phys.\ Rev.\ Lett.\  {\bf 92}, 101602 (2004)
  [arXiv:hep-ph/0308024].
  B.~Gripaios,
  Phys.\ Lett.\  B {\bf 663}, 419 (2008)
  [arXiv:0803.0497 [hep-ph]];
  B.~Gripaios and S.~M.~West,
  Nucl.\ Phys.\  B {\bf 789}, 362 (2008)
  [arXiv:0704.3981 [hep-ph]];
  T.~Flacke, B.~Gripaios, J.~March-Russell and D.~Maybury,
  JHEP {\bf 0701}, 061 (2007)
  [arXiv:hep-ph/0611278].


\bibitem{compaxion}
  J.~E.~Kim,
  Phys.\ Rev.\  D {\bf 31}, 1733 (1985).

\bibitem{compaxion1}
  D.~B.~Kaplan,
  Nucl.\ Phys.\  B {\bf 260}, 215 (1985).

\bibitem{compaxion2}
  L.~Randall,
  Phys.\ Lett.\  B {\bf 284}, 77 (1992).

\bibitem{nosmallgaugecouplings}
  N.~Arkani-Hamed, L.~Motl, A.~Nicolis and C.~Vafa,
  JHEP {\bf 0706}, 060 (2007)
  [arXiv:hep-th/0601001].

\bibitem{fakegaugino}
  C.~Csaki, J.~Heinonen, J.~Hubisz and Y.~Shirman,
  Phys.\ Rev.\  D {\bf 79}, 105016 (2009)
  [arXiv:0901.2933 [hep-ph]].
  
\bibitem{anomalies}
  S.~L.~Adler,
  Phys.\ Rev.\  {\bf 177}, 2426 (1969).
  J.~S.~Bell and R.~Jackiw,
  Nuovo Cim.\  A {\bf 60}, 47 (1969).
  W.~A.~Bardeen,
  Phys.\ Rev.\  {\bf 184}, 1848 (1969).
  G.~'t Hooft,
  Phys.\ Rev.\  D {\bf 14}, 3432 (1976)
  [Erratum-ibid.\  D {\bf 18}, 2199 (1978)].

\bibitem{fujikawa}
  K.~Fujikawa,
  Phys.\ Rev.\ Lett.\  {\bf 42}, 1195 (1979).
  K.~Fujikawa,
  Phys.\ Rev.\  D {\bf 21}, 2848 (1980)
  [Erratum-ibid.\  D {\bf 22}, 1499 (1980)].


\bibitem{nimanom}
  N.~Arkani-Hamed, A.~G.~Cohen and H.~Georgi,
  Phys.\ Lett.\  B {\bf 516}, 395 (2001)
  [arXiv:hep-th/0103135].
  
\bibitem{warpedanom}
  T.~Hirayama and K.~Yoshioka,
  JHEP {\bf 0401}, 032 (2004)
  [arXiv:hep-th/0311233].

\bibitem{PDG}
  C.~Amsler {\it et al.}  [Particle Data Group],
  Phys.\ Lett.\  B {\bf 667}, 1 (2008).

\bibitem{DFSZ}
  M.~Dine, W.~Fischler and M.~Srednicki,
  Phys.\ Lett.\  B {\bf 104}, 199 (1981).
  A.~R.~Zhitnitsky,
  Sov.\ J.\ Nucl.\ Phys.\  {\bf 31}, 260 (1980)
  [Yad.\ Fiz.\  {\bf 31}, 497 (1980)].

\bibitem{symmetrybreaking}
  H.~M.~Georgi, L.~J.~Hall and M.~B.~Wise,
  Nucl.\ Phys.\  B {\bf 192}, 409 (1981).


\bibitem{D0search}
  V.~M.~Abazov {\it et al.}  [D0 Collaboration],
  Phys.\ Rev.\ Lett.\  {\bf 101}, 111802 (2008)
  [arXiv:0806.2223 [hep-ex]].


\bibitem{CDFsearch}
  T.~Aaltonen {\it et al.}  [CDF Collaboration],
  Phys.\ Rev.\  D {\bf 78}, 032015 (2008)
  [arXiv:0804.1043 [hep-ex]].

\bibitem{astroED}
  V.~D.~Barger, T.~Han, C.~Kao and R.~J.~Zhang,
  Phys.\ Lett.\  B {\bf 461}, 34 (1999)
  [arXiv:hep-ph/9905474].

\bibitem{CalcHep}
  A.~Pukhov,
  arXiv:hep-ph/0412191.

\bibitem{gaugefixing}
L.~Randall and M.~D.~Schwartz,
  JHEP {\bf 0111}, 003 (2001)
  [arXiv:hep-th/0108114];
A.~Muck, A.~Pilaftsis and R.~Ruckl,
  Phys.\ Rev.\  D {\bf 65}, 085037 (2002)
  [arXiv:hep-ph/0110391];
 G.~Cacciapaglia, C.~Cs\'aki, C.~Grojean, M.~Reece and J.~Terning,
  Phys.\ Rev.\  D {\bf 72}, 095018 (2005)
  [arXiv:hep-ph/0505001].








\end{thebibliography}
\end{document}